\documentclass{article}

    \PassOptionsToPackage{numbers, compress}{natbib}
 \usepackage[eandd, final]{neurips_2026}

\usepackage[utf8]{inputenc} % allow utf-8 input
\usepackage[T1]{fontenc}    % use 8-bit T1 fonts
\usepackage{hyperref}       % hyperlinks
\hypersetup{hidelinks}
\usepackage{url}            % simple URL typesetting
\usepackage{booktabs}       % professional-quality tables
\usepackage{amsfonts}       % blackboard math symbols
\usepackage{nicefrac}       % compact symbols for 1/2, etc.
\usepackage{microtype}      % microtypography
\usepackage[table]{xcolor}  % colors and table row shading
\usepackage{multirow}
\usepackage{booktabs}    % \toprule, \midrule, \bottomrule
\usepackage{graphicx}    % \resizebox
\usepackage{amssymb}     % \checkmark
\usepackage{array}       % 표 정렬 보조
\usepackage{caption}     % caption formatting 개선
\usepackage{adjustbox}
\usepackage{wrapfig}
\usepackage{tabularx}
\usepackage{float}
\usepackage{caption}
\usepackage{booktabs}
\usepackage{amssymb}

\newsavebox{\promptboxsave}
\newenvironment{promptbox}[1][]{%
  \par\smallskip
  \noindent
  \begin{lrbox}{\promptboxsave}%
  \begin{minipage}{0.92\linewidth}%
  \footnotesize\ttfamily\raggedright
}{%
  \end{minipage}%
  \end{lrbox}%
  \begingroup
  \setlength{\fboxsep}{6pt}%
  \fcolorbox{gray!40}{gray!8}{\usebox{\promptboxsave}}%
  \endgroup
  \par\smallskip
}

\newcommand{\cmark}{\textcolor{green!50!black}{$\checkmark$}}
\newcommand{\xmark}{\textcolor{red!55!black}{$\times$}}
\newcommand{\pmark}{\textcolor{orange!80!black}{$\triangle$}}

\title{VocalCoachBench: Benchmarking Audio-Language Models on Expert Feedback for Singing}

\author{%
  \textbf{Hayeon Bang} \hspace{0.3cm}
  \textbf{Hounsu Kim} \hspace{0.3cm}
  \textbf{Wonil Kim} \hspace{0.3cm}
  \textbf{Juhan Nam} \\[6pt]
  KAIST \\[6pt]
  \texttt{\{hayeonbang,hanshounsu,ianwonilkim,juhan.nam\}@kaist.ac.kr}
}

\begin{document}

\maketitle

\begin{abstract}
Recent audio-language models are increasingly evaluated on recognizing, describing, and reasoning about audio, but expert-facing applications require a different capability: producing feedback that identifies problems and suggests corrective actions grounded in the input. 
We introduce \textbf{VocalCoachBench}, a benchmark for evaluating audio-language models on expert vocal coaching feedback for singing. 
VocalCoachBench contains 515 recordings annotated by 18 professional vocal trainers, yielding 1,064 expert audio-level reviews and 12,216 atomic coaching claims. 
It comprises a same-song subset for controlled comparison and a diverse-song subset for segment-grounded feedback across varied songs and recording conditions. 
To accommodate the open-ended nature of expert feedback, VocalCoachBench separates deterministic structured targets from claim-based assessment of free-form diagnosis and corrective guidance. 
Human annotation analysis shows that expert agreement varies strongly with label granularity, motivating hierarchical structured metrics and claim-based evaluation of open-ended feedback. 
Experiments with 12 recent audio-language models reveal a persistent gap: while the strongest models can compare performances and identify broad issue domains in free-form feedback, the best Top-3 macro-F1 only slightly exceeds a label-prior baseline, and strict diagnosis hit rates remain below 8\%. 
To our knowledge, VocalCoachBench provides the first public testbed for evaluating audio-grounded expert feedback for singing, moving audio-language evaluation beyond description toward analytic feedback.
Data, code, and audio examples are available at \url{https://hayeonbang.github.io/VocalCoachBench}.
\end{abstract}

\section{Introduction}
Audio-language models increasingly understand diverse audio inputs and interact through natural language~\citep{salmonn, qwen2_audio, audio_flamingo}.
Their benchmarks primarily test recognition, description, and question answering~\citep{airbench, mmau, audiobench}.
Expert-facing applications additionally require identifying domain-specific problems and suggesting corrective actions grounded in the input, capabilities that existing benchmarks rarely test.

Vocal coaching requires this form of feedback.
Coaches jointly consider pitch, rhythm, breath support, vocalization, and expression to identify the main issues and provide concrete corrective directions.
Evaluating this ability requires assessing whether feedback identifies salient problems and offers appropriate guidance grounded in the performance.

Prior work has addressed parts of this problem, but has not treated vocal coaching itself as the target of benchmark evaluation.
Automatic singing assessment has long evaluated singing proficiency through acoustic features, perceptual ratings, scores, or rankings~\citep{tsai2012karaoke,gupta2018perceptual,gupta2020leaderboard}, and has more recently expanded toward multi-dimensional perceptual assessment~\citep{vocalverse}.
Separately, datasets and frameworks providing free-text feedback on musical performance have emerged~\citep{neuropiano,crocus,jiang2023expertnovice,llaqo}, but these primarily target piano or instrumental performance.
Moreover, these resources do not formalize feedback quality itself as an evaluation target along the axes of diagnosis, correction, or segment grounding.
VocalCoachBench is motivated by the gap between these two lines of work: the absence of a benchmark that structures and evaluates expert coaching feedback for singing audio.

Building such a benchmark is difficult because expert coaching feedback is inherently open-ended. 
For the same performance, trainers may prioritize different issues or suggest different corrective strategies, making a single scalar score or unique reference response inadequate. 
This challenge is consistent with prior reports of rater variability in music performance assessment~\citep{thompson2003evaluation, wesolowski2015rater} and critiques of non-diagnostic scalar scores in singing assessment~\citep{vocalverse}, and is further supported by our pilot studies in Section~\ref{sec:design_from_pilot}. 
VocalCoachBench therefore separates structured targets, which can be evaluated deterministically, from claim-based open-ended coaching feedback, which accommodates multiple valid expert responses.

We propose \textbf{VocalCoachBench}, which combines constrained labels and rankings for deterministic evaluation with natural-language coaching reviews for claim-based assessment.

% \vspace{-1.0em}
\begin{figure}[t]
  \centering
  \includegraphics[width=\textwidth]{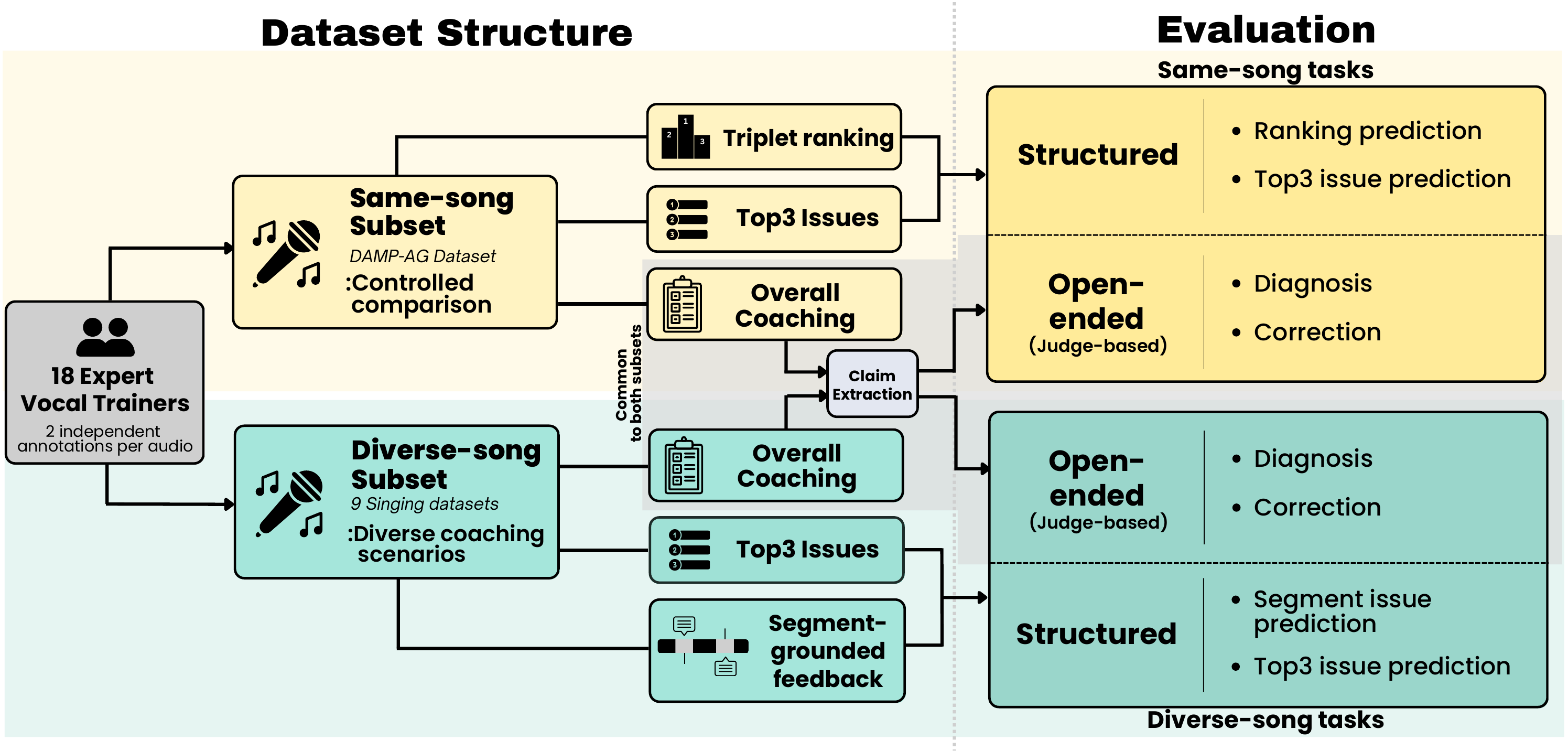}
    \caption{
Overview of VocalCoachBench. Expert vocal trainers annotate two complementary subsets: a same-song subset for controlled comparison, and a diverse-song subset for diverse coaching scenarios and segment-grounded feedback. Both subsets include overall coaching reviews and ranked Top-3 issue labels. Expert feedback is post-processed into atomic coaching claims, which support structured and open-ended evaluation.}
  \label{fig:overview}
\end{figure}

Eighteen expert vocal trainers provide ranked Top-3 issue labels and coaching reviews covering problems, corrective directions, and strengths. A same-song subset supports controlled triplet rankings with rationales, while a diverse-song subset adds segment-grounded feedback with time spans.

% VocalCoachBench combines a same-song subset for controlled comparison with a diverse-song subset for diverse coaching scenarios.
% Expert trainers provide ranked issue labels, natural-language coaching feedback, triplet comparisons, and segment-grounded feedback. These annotations support deterministic structured tasks and claim-based open-ended evaluation.

% VocalCoachBench annotations are built on long-form coaching feedback authored by 18 expert vocal trainers.
% Each expert submission includes a ranked Top-3 selection of vocal issue categories and an overall coaching review written in natural language, with instructions to discuss salient problems, corrective directions, and strengths.
% In addition, the same-song subset includes controlled triplet rankings together with ranking rationales, and the diverse-song subset includes segment-grounded feedback with time spans.
% This design preserves the richness of expert feedback while enabling both within-condition comparison and evaluation across diverse coaching scenarios.

The \emph{structured evaluation suite} comprises same-song triplet ranking, Top-3 issue prediction, and expert-localized segment issue classification.
The \emph{open-ended coaching suite} evaluates generated feedback against expert atomic claims, separating diagnosis coverage from correction validity and accommodating multiple expert perspectives.

Our contributions are summarized as follows:
\vspace{-5pt}
\begin{enumerate}
\item \textbf{Expert-annotated vocal coaching benchmark.}
We introduce VocalCoachBench, to our knowledge, the first publicly released benchmark annotations for singing audio that include expert free-text coaching feedback, structured vocal issue labels, controlled rankings, and segment-grounded vocal issue annotations.
\vspace{-3pt}
\item \textbf{Benchmark design for partially aligned expert feedback.}
We introduce a design that combines controlled same-song comparison, diverse-song coaching, and claim-based evaluation to handle partially aligned expert feedback without reducing it to a single score or reference response.
\vspace{-3pt}
\item \textbf{Evaluation suite and model benchmark.}
We define deterministic structured tasks and claim-based open-ended evaluations for diagnosis and correction quality, and benchmark 12 recent audio-language models on vocal coaching.
\end{enumerate}
% \vspace{-4pt}

\begin{table}[t]
\centering
\caption{Comparison with representative resources for singing assessment, music performance feedback, and expert skill feedback.}
\label{tab:dataset_comparison_main}
\footnotesize
\setlength{\tabcolsep}{3.2pt}
\renewcommand{\arraystretch}{0.92}
\resizebox{\linewidth}{!}{%
\begin{tabular}{llcccc}
\toprule
\textbf{Work} & \textbf{Main target} &
\textbf{Structured Eval} & \textbf{Expert FB} & \textbf{Diag/Corr Eval} & \textbf{Segment-level FB} \\
\midrule

\rowcolor{gray!8}
\multicolumn{6}{l}{\textit{Singing assessment}} \\
SingEval~\citep{gupta2020leaderboard}
& Quality score/rank
& \cmark & \xmark & \xmark & \xmark \\
% VocalVerse~\citep{vocalverse}
% & Multidimensional assessment
% & \cmark & \pmark & \xmark & \xmark \\

\midrule
\rowcolor{gray!8}
\multicolumn{6}{l}{\textit{Music performance assessment and feedback}} \\
PercePiano~\citep{percepiano}
& Perceptual evaluation
& \cmark & \xmark & \xmark & \xmark \\
NeuroPiano~\citep{neuropiano}
& Rating + feedback
& \cmark & \cmark & \xmark & \xmark \\
% LLaQo \citep{llaqo}
% & Query-based coaching 
% & \cmark & \cmark & \xmark & \xmark \\
CROCUS~\citep{crocus}
& Formative critiques
& \xmark & \cmark & \xmark & \pmark \\
Expert-Novice~\citep{jiang2023expertnovice}
& Rating + feedback
& \cmark & \cmark & \xmark & \xmark \\

\midrule
\rowcolor{gray!8}
\multicolumn{6}{l}{\textit{Embodied skill feedback}} \\
VidDiffBench~\citep{viddiffbench}
& Action differencing
& \xmark & \cmark & \xmark & \cmark \\
% ExAct~\citep{exact}
% & Expert action QA
% & \cmark & \pmark & \xmark & \xmark \\
Ego-Exo4D~\citep{egoexo}
& Proficiency/commentary
& \pmark & \cmark & \xmark & \cmark \\

\midrule
\rowcolor{gray!8}
\multicolumn{6}{l}{\textit{Ours}} \\
\textbf{VocalCoachBench}
& \textbf{Vocal coaching}
& \cmark & \cmark & \cmark & \cmark \\
\bottomrule
\end{tabular}%
}
\vspace{0.25em}

\begin{minipage}{0.95\linewidth}
\scriptsize
\textit{Notes.}
Structured Eval = constrained targets such as scores, rankings, classifications, or QA.
Diag/Corr Eval = whether diagnosis and correction are explicit evaluation targets.
Segment-level FB = timestamped or otherwise localized feedback.

\cmark = provided; \pmark = partial, implicit, or not directly evaluated; \xmark = absent.
\end{minipage}
\end{table}
\section{Related work}
\label{sec:related_work}
\vspace{-4pt}
% \subsection{Audio-language models and audio benchmarks}
% Recent audio-language models such as SALMONN, Qwen-Audio/Qwen2-Audio, and Audio Flamingo extend large language models to speech, music, and environmental audio, enabling audio captioning, question answering, instruction following, and dialogue \citep{salmonn, qwen_audio, qwen2_audio, audio_flamingo}. 
% Corresponding benchmarks such as AIR-Bench, MMAU, and AudioBench evaluate general audio understanding, reasoning, and instruction following \citep{airbench,mmau,audiobench}. 
% Music and audio captioning datasets further emphasize descriptive language generation for audio events or musical content, but they do not evaluate whether generated text identifies domain-specific performance problems or provides corrective guidance.
% These works establish broad audio-language capabilities, but they do not directly evaluate domain-specific expert feedback grounded in audio. 
% VocalCoachBench addresses this gap in the setting of vocal coaching. 
\subsection{Audio-language models and audio benchmarks}

Recent audio-language models such as SALMONN, Qwen-Audio/Qwen2-Audio, and Audio Flamingo extend large language models to speech, music, and environmental audio, enabling captioning, question answering, instruction following, and dialogue~\citep{salmonn, qwen_audio, qwen2_audio, audio_flamingo}. 
Benchmarks and datasets for audio understanding — ranging from general evaluation suites such as AIR-Bench, MMAU, and AudioBench~\citep{airbench,mmau,audiobench} to music description datasets such as LP-MusicCaps and SongDescriber~\citep{lpmusiccaps,songdescriber} — focus on recognition, description, or captioning, and do not evaluate whether models can identify domain-specific performance problems or provide corrective guidance grounded in audio.

% \subsection{Automatic singing assessment} 
% Automatic singing assessment has long studied how to evaluate singing proficiency from acoustic and perceptual cues. 
% Early work used pitch-interval accuracy and vibrato for reference-independent skill assessment~\citep{nakano2006singing}, while karaoke scoring systems combined pitch, volume, and rhythm~\citep{tsai2012karaoke}. 
% Later work modeled richer perceptual dimensions such as intonation, rhythm, vibrato, timbre, and pronunciation~\citep{gupta2018perceptual}, and recent work has moved toward reference-free quality estimation, rank ordering, timbre-guided evaluation, and multidimensional perceptual assessment~\citep{gupta2020leaderboard,gupta2020rankordering,tg_critic,vocalverse}. 
% At the same time, reusable public datasets with expert annotations remain limited in singing assessment, and existing resources are often small, platform-specific, non-public, or focused on scores rather than coaching feedback.
% However, this line of work primarily evaluates singing quality, perceptual dimensions, or ranking validity. It does not benchmark whether a model can identify coach-relevant vocal problems and generate corrective feedback grounded in the audio.

\subsection{Automatic singing assessment} 

Automatic singing assessment has long studied singing proficiency from acoustic and perceptual cues. 
Early work used pitch-interval accuracy and vibrato for reference-independent skill assessment~\citep{nakano2006singing}, while karaoke scoring systems combined pitch, volume, and rhythm~\citep{tsai2012karaoke}. 
Later work modeled perceptual dimensions such as intonation, rhythm, vibrato, timbre, and pronunciation~\citep{gupta2018perceptual}, and recent work has moved toward reference-free quality estimation, rank ordering, timbre-guided evaluation, and multidimensional perceptual assessment~\citep{gupta2020leaderboard,gupta2020rankordering,tg_critic,vocalverse}. 
However, singing assessment primarily focuses on evaluating vocal quality, perceptual dimensions, scores, or ranking validity. In addition, reusable public datasets for singing assessment remain limited, as existing resources are often small, platform-specific, non-public, or focused on scores rather than coaching feedback.

% \subsection{Expert feedback for music performance and embodied skills}
% A separate line of work has studied expert annotation and natural-language feedback for music performance. 
% NeuroPiano, CROCUS, and Expert-Novice include expert ratings, critiques, or formative feedback for piano and instrumental performance~\citep{neuropiano,crocus,jiang2023expertnovice}, and LLaQo explores query-based coaching for expressive piano performance using such resources~\citep{llaqo}. 
% These works are closest in spirit to VocalCoachBench, but remain piano- or instrument-centered and do not evaluate feedback quality as vocal diagnosis, correction, or segment-grounded coaching.

% Beyond music, Ego-Exo4D and VidDiffBench provide expert commentary, proficiency annotations, or skill-relevant differences for video-based embodied activities~\citep{egoexo,viddiffbench}; ExAct further converts Ego-Exo4D commentary into expert-level video QA~\citep{exact}. 
% They show broader interest in expert feedback for human performance, but are video-centric and do not address audio-grounded vocal coaching. 

% Table~\ref{tab:dataset_comparison_main} summarizes representative resources.
% Prior datasets cover singing assessment, expert feedback, or localized commentary, but do not explicitly evaluate diagnosis and correction for vocal coaching.

\subsection{Expert feedback for music performance and skill-based activities}
A separate line of work studies expert annotation and natural-language feedback for music performance. 
NeuroPiano, CROCUS, and Expert-Novice collect expert ratings, critiques, or formative feedback for piano and instrumental performance~\citep{neuropiano,crocus,jiang2023expertnovice}, while LLaQo turns performance-understanding resources into query-response coaching data~\citep{llaqo}.
Beyond music, recent embodied-skill benchmarks evaluate expert-level performance understanding through proficiency cues, improvement-related commentary, action differences, or expert-derived QA~\citep{egoexo,viddiffbench,exact}.
% ExAct further converts Ego-Exo4D commentary into expert-level video QA~\citep{exact}. 
VocalCoachBench complements these efforts by moving from collecting or matching expert feedback to evaluating the internal quality of generated feedback: whether its diagnosis is expert-aligned and whether its correction is appropriate to the diagnosed issue.

Table~\ref{tab:dataset_comparison_main} summarizes representative resources. 
Prior datasets cover singing assessment, expert feedback, or localized commentary, but do not explicitly evaluate diagnosis and correction for vocal coaching.

\section{VocalCoachBench construction}
\label{sec:construction}

We structure expert feedback into evaluable targets reflecting coaching practice: identifying the main problems, explaining them in pedagogical terms, and proposing corrective directions.

\subsection{Design motivation from pilot studies}
\label{sec:design_from_pilot}

Following prior work in singing assessment, our initial design considered including scalar ratings.
However, recent singing assessment work has questioned whether non-diagnostic scalar scores can adequately represent complex vocal performance judgments~\citep{vocalverse}.
Across two pilot studies, we observed the same limitation in the coaching setting: absolute scores proved to be too compressed a format for expressing vocal coaching judgments.
Based on this observation, the second pilot restricted annotators to four pop vocal experts to reduce domain variation, and applied a pre-agreed rating scale to a shared set of same-song recordings.

Despite this constraint, scalar ratings showed limited agreement, with an overall Krippendorff's $\alpha$ of approximately 0.323 and aspect-level $\alpha$ values ranging from approximately 0.171 to 0.393.
Higher within-one-point agreement suggested that experts often made nearby judgments using different thresholds.
In follow-up discussions, trainers described nonuniform scoring practices and different severity thresholds, suggesting that scalar-score disagreement partly reflects the elicitation format.
% We therefore removed rating scores from the final dataset and redesigned the annotation protocol around concrete coaching actions.
% VocalCoachBench decomposes expert coaching into five evaluable components: issue identification, diagnosis-side issue claims, corrective guidance, relative ranking, and segment-grounded feedback.
We therefore removed rating scores from the final dataset and redesigned the annotation protocol around concrete coaching actions: issue identification, diagnosis claims, corrective guidance, controlled comparison, and segment-grounded feedback.

\subsection{Audio sources and dual-subset design}
\label{sec:audio_sources}

This redesign also shapes how we construct the benchmark.
Evaluating relative ranking requires that different performances be comparable under conditions as similar as possible, whereas real vocal coaching addresses problems that arise across diverse songs, styles, and vocal registers.
Because no single subset can satisfy both comparability and musical diversity, VocalCoachBench is organized into a same-song subset and a diverse-song subset.

\textbf{The same-song subset} is built on \emph{Amazing Grace} recordings from DAMP-S-AG \citep{damp_sag}.
Because all performers sing the same song, differences in lyrics, melody, difficulty and song structure are controlled, making relative comparison more meaningful than in a diverse-song setting.
To sample performances across different proficiency levels, we used an automatically computed pitch-accuracy proxy to group candidates into high, middle, and low tiers, and then sampled 207 unique recordings in an approximately 1:2:1 ratio.
The proxy is used only for this sampling.

\textbf{The diverse-song subset} provides the musical diversity that the same-song subset lacks.
We collected English solo singing audio from nine publicly available singing datasets: PopBuTFy \citep{popbutfy}, N20EMv2 \citep{n20emv2}, LM-SSD \citep{LM-SSD}, NUS-48E \citep{nus48e}, MedleyDB \citep{medleydb}, MedleyVox~\citep{medleyvox}, MRSSing \citep{mrssing}, URSing \citep{ursing}, and GTSinger \citep{gtsinger}.
When singer information was available, we capped the number of recordings per singer at four to avoid over-representing any individual.
The final diverse-song subset consists of 308 unique recordings.

Two authors screened all candidate recordings and removed samples with very low audio quality, non-singing speech-like vocals, multiple simultaneous singers, non-English lyrics, or accompaniment that obscured the vocal. For long recordings, we selected annotation-length excerpts using natural pauses or repeated-section boundaries.

\subsection{Expert annotation protocol}
\label{sec:annotation_protocol}
%돈 
Annotation was carried out by 18 professional vocal trainers with formal vocal-performance training and at least three years of coaching experience.  
Six annotators worked on the same-song subset and twelve on the diverse-song subset. 
Of the 515 recordings, 504 received two independent expert reviews, nine shared audit recordings received six, and two received one.

\begin{wraptable}{r}{0.4\textwidth}
% \vspace{-10pt}
\centering
\caption{
VocalCoachBench dataset summary.
}
\label{tab:dataset_summary}
\small
\setlength{\tabcolsep}{5pt}
\renewcommand{\arraystretch}{1.12}
\begin{tabular}{@{}lrrr@{}}
\toprule
\textbf{Dataset field} & \textbf{Same} & \textbf{Diverse} & \textbf{Total} \\
\midrule
\textbf{Audio}  & 207   & 308   & \textbf{515} \\
\textbf{Reviews}  & 450   & 614   & \textbf{1,064} \\
\textbf{Blocks} & 1,397 & 3,109 & \textbf{4,506} \\
\textbf{Claims} & 3,972 & 8,244 & \textbf{12,216} \\
\textbf{Diagnosis}  & 2,090 & 4,072 & \textbf{6,162} \\
\textbf{Correction}  & 1,642 & 2,951 & \textbf{4,593} \\
\textbf{Strength}   & 240   & 1,221 & \textbf{1,461} \\
\bottomrule
\end{tabular}
\vspace{-5pt}
\end{wraptable}

The annotation interface was designed to mirror real coaching situations.
Annotators listened to the full recording and wrote a natural-language coaching review addressed to the student.
Annotators were instructed to include the main problems, corresponding corrective directions, and strengths.
To support later analysis and evaluation, annotators wrote their feedback in semantically coherent feedback blocks and tagged each block with one or more vocal issue categories.

The final taxonomy consists of seven fine-grained issue categories under three parent categories: 
\textit{Technical Production} covers \textit{breath support} (use and control of breath), \textit{vocalization} (phonation, resonance, register, and voice placement), and \textit{technique} (specialized skills such as vibrato and ornamentation); 
\textit{Musical Accuracy} covers \textit{pitch} (note accuracy and stability) and \textit{rhythm} (beat and timing); and 
\textit{Delivery} covers \textit{diction} (lyric intelligibility and pronunciation) and \textit{expression} (musical interpretation and emotional delivery). 
The full annotator guidance is given in Appendix~\ref{app:taxonomy_guidance}.

The same-song subset additionally includes triplet ranking: after the overall review and ranked Top-3 issue selection, annotators ranked three performances and wrote a ranking rationale.
Each of three trainer pairs ranked 75 triplets built from 75 recordings: 66 assigned only to that pair and the nine shared audit recordings.
Triplets were formed by length-three sliding windows over a tier-stratified circular sequence, so each recording appears in three of its pair's triplets; the three triplets of shared recordings were ranked by all six trainers.
The diverse-song subset additionally includes segment-grounded feedback: annotators selected main problem regions on a waveform display and wrote segment-grounded comments.
Segments were not collected for the same-song subset, where shared melody, lyrics, and phrase structure could lead trainers to mark the same difficult passages across performers, so that song position rather than recording-specific evidence would drive segment evaluation.

\subsection{Quality control and post-processing}
\label{sec:quality_control}
The original annotations were written in Korean. 
We used an LLM-assisted pipeline to decompose audio-level feedback blocks into atomic \emph{diagnosis}, \emph{correction}, and \emph{strength} claims, link each correction to its corresponding issue, and translate the claims into English. The pipeline works from the expert-written text and its issue tags alone; the prompt prohibits unsupported additions and uses the tags only as hints. We spot-checked the translations for vocal terminology and coaching intent. In a 100-block audit, we observed no omitted or hallucinated claims and only minor extraction errors; details are provided in Appendix~\ref{app:claim_audit}.

\begin{figure*}[t!]
\centering
\includegraphics[width=\textwidth]{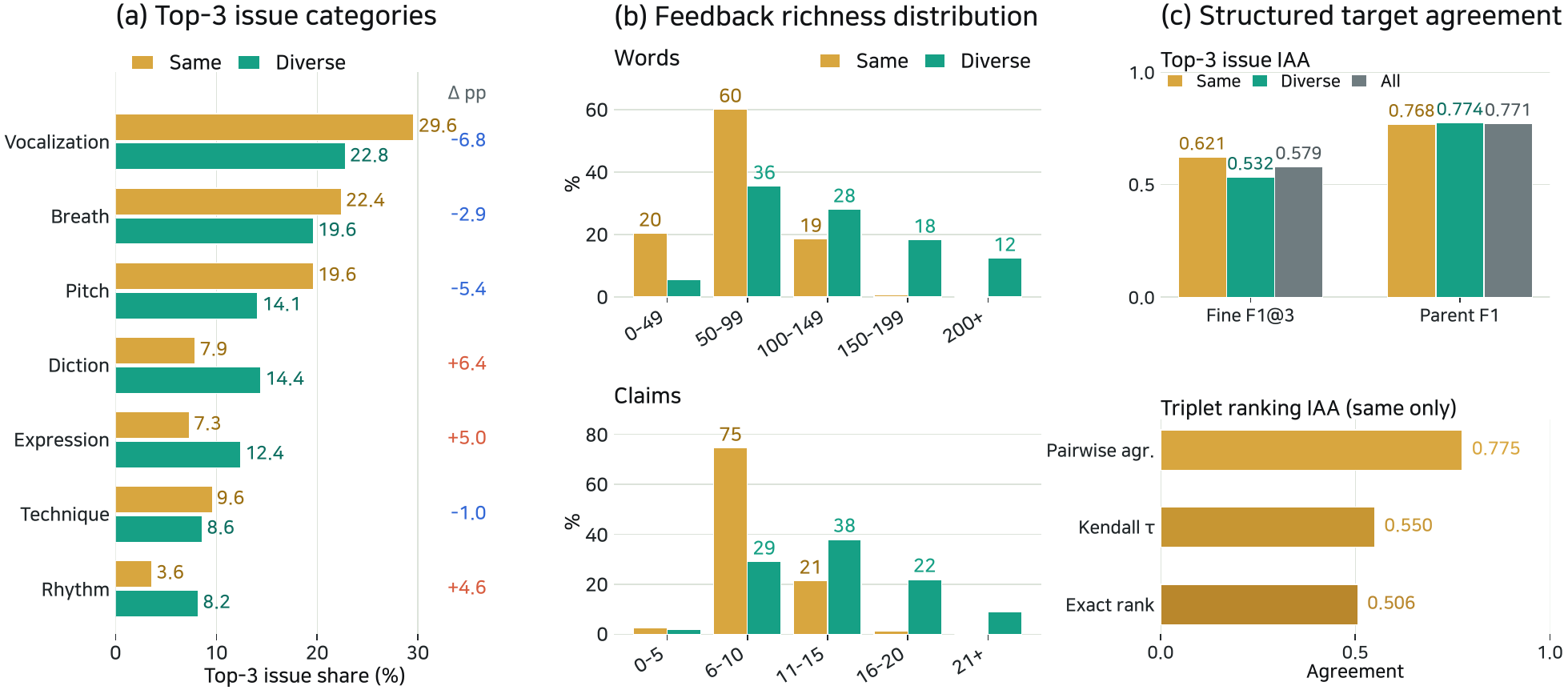}
\caption{
Annotation statistics and agreement in VocalCoachBench. (a) Top-3 issue distributions across subsets, with percentage-point shifts shown on the right. (b) Feedback richness in words and atomic claims per annotator-audio submission. (c) Expert agreement on structured targets, including parent-level Top-3 agreement, fine-label agreement, and same-song triplet rankings.
}
\label{fig:dataset_analysis}
\end{figure*}
\subsection{Dataset statistics and annotation behavior}
\label{sec:dataset_statistics}
VocalCoachBench contains 515 recordings, 1,064 expert audio-level reviews, 4,506 feedback blocks, and 12,216 atomic claims (Table~\ref{tab:dataset_summary}).
The median review contains four feedback blocks and ten atomic claims; diverse-song reviews contain more audio-level feedback than same-song reviews, with medians of five versus three blocks and thirteen versus eight claims.

Figure~\ref{fig:dataset_analysis} summarizes annotation behavior. The same-song and diverse-song subsets show different Top-3 issue profiles: same-song annotations concentrate more on vocalization, breath support, and pitch, whereas diverse-song annotations include relatively more diction, expression, and rhythm. This supports the dual-subset design.
Feedback blocks are also multi-factorial: 68.6\% are tagged with two or more issue categories.

Human agreement varies by representation level. Across the 513 recordings with at least two expert reviews, recording-balanced expert-pair agreement is 0.573 for fine-level F1@3 and 0.777 for parent-level F1, compared with 0.284 for fine-level Top-1 agreement. Category-level agreement, measured by one-vs-rest F1 between expert Top-3 sets, is highest for vocalization (0.818) and lowest for rhythm (0.268) and technique (0.256; Appendix~\ref{app:category_analysis}).
Across 261 expert-order pairs from 219 annotated same-song triplets, pairwise agreement is 77.5\% and mean Kendall's $\tau$ is 0.550.
Segment annotations are temporally open-ended, but the frozen evaluation set contains 262 consensus segments formed from overlapping expert spans with IoU$\geq$0.3 and a shared issue tag.

These patterns motivate our evaluation design: constrained rankings, Top-3 issue prediction, and expert-localized segment issue classification are evaluated deterministically, while open-ended diagnosis and correction are evaluated through atomic claims.

\section{Evaluation suite}
\label{sec:eval_suite}
VocalCoachBench combines a structured suite for performance ranking and issue identification with a claim-based suite for open-ended coaching.
All evaluations use the 515 released recordings.
We score the first stored response of each model: missing or unparseable predictions remain in the evaluation population rather than being replaced by successful retries.
% Results artifact: rebuttal_experiments/20261026_final_leaderboard/leaderboard.json
% Dataset version: canonical515_physical_audio_1064refs
% Policy: strict_all_attempts; stored first responses; recording-balanced expert references;
% corrected515 format-reported Claims with fixed slots.

\subsection{Structured evaluation suite}
\label{sec:structured_tasks}
\paragraph{Triplet ranking}
Triplet ranking evaluates relative preferences among three same-song performances.
We obtain three direct pairwise decisions per triplet and compare each with both experts' preferences on the 189 triplets whose two expert rankings agree on at least two of three pairs, yielding 567 model decisions and 1,134 expert-reference comparisons.
We report pairwise accuracy, exact-order accuracy, and cycle/tie rate, which includes tied win counts caused by missing decisions. We also report the evaluator's pairwise concordance $\tau_{\rm pair}$: the mean of $+1$ for agreement, $-1$ for disagreement, and $0$ for a missing decision. It equals Kendall's $\tau$ for complete, transitive rankings.
We report score-derived pairwise accuracy from single-audio quality scores on the same comparisons; tied or missing scores receive no credit.

\paragraph{Top-3 issue prediction}
The model selects and ranks three labels from the seven-category taxonomy (Section~\ref{sec:annotation_protocol}).
Evaluation covers all 515 recordings and 1,064 expert annotations.
Each recording has equal total weight, divided among its expert references.
The primary metric is macro-F1 across the seven one-vs-rest classes. We also report F1@3 (Top-3 set overlap divided by three), Top-1 accuracy, nDCG@3 (linear relevance gains of 3, 2, and 1), and parent-level F1 (set F1 over the three parent categories). All overlap metrics average expert references within each recording before averaging recordings.
A non-audio majority baseline always predicts the three most prevalent labels: vocalization, breath support, and pitch.
For reference, an F1@3 of 50\% corresponds to 1.5 shared labels on average, and selecting three labels uniformly at random yields an expected F1@3 of 42.86\%.

\paragraph{Expert-localized segment issue classification}
This task evaluates local issue recognition on 262 expert-consensus clips.
The model receives the expert-cropped audio and predicts one main issue; event discovery and boundary accuracy are not evaluated.
We report seven-class macro-F1, treating all shared reference labels as positives, and any-match accuracy, which credits a prediction matching any shared label.

\subsection{Open-ended coaching evaluation}
\label{sec:open_ended_eval}
Open-ended evaluation separates diagnosis coverage from prescription validity and their end-to-end alignment.
A text-only judge evaluates extracted model claims using expert diagnosis claims and vocal-pedagogy criteria through rubric-based semantic comparison rather than lexical or embedding similarity; strength claims are not scored.
We retain up to five model diagnoses and five prescriptions per recording.
Unparseable coaching outputs contribute no claims, while all reference claims and prescription slots remain in the denominators.

\paragraph{Diagnosis coverage}
For each of 6,162 expert diagnosis claims, the judge assigns \emph{strict hit} (the same concrete problem), \emph{coarse hit} (the same broader issue without its specific details), \emph{miss}, or \emph{contradict}.
Strict hit rate and hit rate divide strict hits and strict-or-coarse hits by all 6,162 reference claims.
These are claim-weighted coverage metrics, unlike the recording-balanced Top-3 evaluation.

\paragraph{Correction appropriateness}
The judge assesses each model prescription against its target model diagnosis, with expert issue claims as context, rather than requiring agreement with a particular expert correction.
Labels are \emph{valid}, \emph{weak}, \emph{invalid}, and \emph{vocally inappropriate}, reflecting specificity, issue alignment, actionability, and vocal safety.
Prescription validity is the number of valid prescriptions divided by 2,575 fixed slots ($515\times5$); omitted slots receive no credit.
The end-to-end rate uses the same denominator and credits only valid prescriptions whose target model diagnosis supports at least one expert diagnosis claim through a strict or coarse hit.

\begin{table}[t]
\centering
\caption{
Structured evaluation. All values are percentages except $\tau_{\rm pair}$.
Direct and Score are pairwise accuracies from audio comparisons and single-audio scores (1,134 expert-reference comparisons). Exact uses 378 expert-order comparisons; Cycle/tie uses 189 triplets.
$\tau_{\rm pair}$ is mean pairwise concordance, with missing decisions contributing zero.
Top-3 metrics use 515 recordings and 1,064 recording-balanced references; Segment metrics use 262 clips. Match is any-match accuracy.
Closed models use Main prompts via APIs; bold marks the group maximum for accuracy, F1, nDCG, and concordance.
$^{\dagger}$Direct uses one concatenated waveform (A, 1\,s silence, B). -- denotes not evaluated.}
\label{tab:structured_protocol_results}
\scriptsize
\setlength{\tabcolsep}{1.85pt}
\renewcommand{\arraystretch}{1.10}
\begin{adjustbox}{max width=\textwidth}
\begin{tabular}{@{}ll*{12}{c}@{}}
\toprule
\multirow{2}{*}{\textbf{Protocol}} &
\multirow{2}{*}{\textbf{Model}} &
\multicolumn{5}{c}{\textbf{Triplet ranking}} &
\multicolumn{5}{c}{\textbf{Top-3 issue prediction}} &
\multicolumn{2}{c}{\textbf{Segment}} \\
\cmidrule(lr){3-7}\cmidrule(lr){8-12}\cmidrule(lr){13-14}
& & \textbf{Direct} & \textbf{Score} & \textbf{$\tau_{\rm pair}$} & \textbf{Exact} & \textbf{\shortstack{Cycle/tie\\$\downarrow$}} &
\textbf{\shortstack{Macro-\\F1}} & \textbf{F1@3} & \textbf{Top-1} & \textbf{\shortstack{nDCG\\@3}} & \textbf{\shortstack{Parent\\F1}} &
\textbf{\shortstack{Macro-\\F1}} & \textbf{Match} \\
\midrule
\rowcolor{gray!8}
\multicolumn{2}{@{}l}{Label-prior baseline (majority, no audio)} & -- & -- & -- & -- & -- & 32.72 & 62.49 & 34.82 & 63.61 & 74.56 & -- & -- \\
\midrule
\multirow{6}{*}{\textsc{Main}} & R1-AQA~\citep{r1aqa} & 50.53 & 0.00 & 0.011 & 13.76 & 26.46 & 21.43 & 34.50 & 13.95 & 30.94 & 55.70 & 7.00 & 32.44 \\
 & Qwen2.5-Omni~\citep{qwen25omni} & \textbf{68.69} & 4.41 & \textbf{0.374} & \textbf{34.13} & 6.35 & 31.07 & 38.33 & 10.91 & 32.84 & 71.39 & 7.67 & 30.92 \\
 & Kimi-Audio~\citep{kimi_audio} & 56.00 & 33.51 & 0.123 & 17.72 & 26.98 & 28.25 & 33.82 & 4.47 & 26.52 & \textbf{78.94} & 13.63 & 29.01 \\
 & MiMo-Audio~\citep{mimo_audio} & 64.46 & \textbf{36.95} & 0.289 & 28.04 & 2.65 & 30.15 & \textbf{42.55} & 15.37 & 38.75 & 78.73 & 8.23 & \textbf{40.46} \\
 & Qwen3-Omni~\citep{qwen3omni} & 64.99 & 13.58 & 0.300 & 27.78 & 16.93 & \textbf{32.32} & 40.88 & 14.40 & 36.74 & 73.22 & \textbf{14.20} & 35.50 \\
 & Fun-Audio-Chat~\citep{fun_audio_llm} & 48.77 & 31.57 & 0.007 & 12.43 & 26.46 & 29.71 & 41.38 & \textbf{33.66} & \textbf{45.57} & 77.32 & 9.11 & 33.59 \\
\midrule
\multirow{4}{*}{\textsc{Closed}} & GPT-Audio 1.5~\citep{gpt_audio} & 71.69 & \textbf{43.74} & 0.434 & 39.15 & 2.12 & \textbf{33.46} & 39.12 & \textbf{14.05} & 34.81 & \textbf{77.83} & 8.09 & 30.15 \\
 & Gemini 3 Flash Preview~\citep{gemini25flash} & 71.16 & 37.65 & 0.423 & 39.95 & 6.88 & 33.40 & \textbf{49.84} & 13.69 & \textbf{44.41} & 74.64 & 12.78 & \textbf{35.50} \\
 & Qwen3.5-Omni Flash~\citep{qwen35omni} & 72.40 & 28.13 & 0.448 & 38.62 & 4.23 & 31.61 & 44.54 & \textbf{14.05} & 39.84 & 74.26 & \textbf{14.65} & 32.82 \\
 & Qwen3.5-Omni Plus~\citep{qwen35omni} & \textbf{74.51} & 40.92 & \textbf{0.490} & \textbf{43.39} & 2.12 & 21.53 & 31.89 & 13.95 & 29.10 & 74.13 & 8.75 & 32.06 \\
\midrule
\multirow{2}{*}{\textsc{Simple}} & Audio Flamingo 3~\citep{audio_flamingo3}$^{\dagger}$ & 49.03 & 0.00 & \textbf{$-$0.009} & 12.17 & 29.10 & 23.28 & 33.67 & \textbf{13.85} & 30.31 & 56.93 & \textbf{13.53} & \textbf{37.40} \\
 & Music Flamingo~\citep{music_flamingo}$^{\dagger}$ & \textbf{49.47} & \textbf{28.40} & $-$0.011 & \textbf{12.70} & 26.46 & \textbf{33.32} & \textbf{43.46} & 13.46 & \textbf{37.87} & \textbf{74.42} & 3.82 & 14.50 \\
\bottomrule
\end{tabular}
\end{adjustbox}
\end{table}

\paragraph{Judge validation}
\label{sec:judge_validation}
We validate the GPT-OSS-120B judge with label-stratified human audits (Appendix~\ref{app:judge_validation}).
Two authors annotate 100 diagnosis cases, yielding 193 usable ratings and 93 jointly labeled cases.
Judge--author agreement is 83.4\% (Cohen's $\kappa=0.779$), and author--author agreement is 80.6\% ($\kappa=0.737$).
A professional vocal coach reviews 100 prescription cases; the 72 eligible \emph{valid}, \emph{weak}, and \emph{invalid} decisions yield 79.2\% agreement ($\kappa=0.687$).
The audited decisions come from an earlier evaluation run; decisions on the final 515-recording inputs have not been re-audited.
On a sample of earlier judgments, a paraphrased judge prompt leaves most decisions unchanged, whereas a second judge model changes strict correction validity and the top Main model; strict validity should therefore be read as judge-dependent, together with the \emph{weak} rate (Appendix~\ref{app:judge_sensitivity}).

\section{Experiments}
\label{sec:experiments}

\subsection{Models and protocol routing}
\label{sec:models}
We include audio-language models that accept raw audio input, can generate open-ended diagnosis and correction without benchmark-specific fine-tuning, and are reproducibly accessible through released checkpoints or public APIs; discriminative singing-assessment systems, which do not generate coaching feedback, fall outside this interface.
We benchmark 12 audio-language models (Table~\ref{tab:structured_protocol_results}): six open-weight systems and four closed-source systems queried through public APIs use Main prompts; Audio Flamingo 3 and Music Flamingo use Simple prompts because they do not reliably follow long schema-constrained prompts.

\subsection{Prompting protocols and inference setup}
\label{sec:prompting_inference}
The Main protocol requests task-specific structured outputs for coaching claims, Top-3 issues, and segment classification.
The Simple protocol uses short natural-language prompts, with a separate prompt for single-audio quality scores.
Direct ranking uses pairwise audio comparisons; Audio Flamingo 3 and Music Flamingo receive concatenated audio with interval instructions.
Model-specific decoding settings and prompts are provided in Appendix~\ref{app:prompts}.

\paragraph{Output parsing and normalization}
Top-3 and Segment outputs are mapped to canonical labels using frozen deterministic rules that read structured fields and recognize label aliases in text.
Quality scores and pairwise winners follow task-specific parsers, with ties, missing scores, and unparseable decisions receiving no ranking credit.
Main coaching claims are parsed from structured responses.
Simple coaching reviews use text-only GPT-OSS-120B extractions, which may identify implied improvement targets; their end-to-end evaluation retains the original positional target-linking rule when explicit links are absent.
Raw responses and extraction provenance are retained.

\section{Results}
Simple-protocol models are grouped separately in Tables~\ref{tab:structured_protocol_results} and~\ref{tab:open_ended_coaching_results} because their prompts and extraction procedures differ from Main.

\vspace{-6pt}
\paragraph{Structured tasks}
Direct comparison yields higher pairwise accuracy than score-derived ranking for all 12 models evaluated on both tasks.
Qwen2.5-Omni reaches 68.69\% Direct accuracy among open-weight Main models, and Qwen3.5-Omni Plus reaches 74.51\% among closed-source models; the highest score-derived accuracy is 43.74\%.
Audio Flamingo 3 and Music Flamingo reach 49.03\% and 49.47\%, respectively, under the concatenated-audio variant.

Fine-grained issue identification remains difficult.
GPT-Audio 1.5 and Gemini 3 Flash Preview attain Top-3 macro-F1 scores of 33.46\% and 33.40\%, only slightly above the 32.72\% majority baseline, and no model exceeds it significantly under a paired recording bootstrap (Appendix~\ref{app:category_analysis}); all six open-weight Main models remain below this baseline.
Music Flamingo reaches 33.32\% under Simple prompts. All models remain below the majority baseline on F1@3 (62.49\%), whose frequent-label overlap complements the class-balanced macro-F1 comparison.
Segment macro-F1 peaks at 14.65\%; MiMo-Audio's 40.46\% Match accuracy results from predicting vocalization for all 262 segments, with macro-F1 of only 8.23\%.
Thus, frequent-label predictions can yield substantial overlap without discriminating fine-grained issues.
Category-level results show model-specific label templates rather than uniform over-prediction of frequent labels: six models never predict expression, whereas six include rhythm, the least frequent expert label, for more than 90\% of recordings (Appendix~\ref{app:category_analysis}).
Supervised heads on frozen audio representations nonetheless exceed metadata-only controls in Top-3 macro-F1 (48.52\% vs.\ 41.74\%), indicating learnable recording-dependent signal (Appendix~\ref{app:supervised_controls}).

\paragraph{Open-ended coaching}
Strict diagnosis hit rates reach at most 7.53\%, while strict-or-coarse coverage reaches 63.11\%, indicating a gap between broad issue coverage and precise expert-supported diagnosis.
Qwen3.5-Omni Plus achieves 96.85\% prescription validity but 63.96\% end-to-end alignment, both measured over 2,575 fixed prescription slots.
Plausible corrective advice therefore need not be linked to an expert-supported diagnosis.
For the closed-source models, validity among emitted prescriptions differs by at most 2.05 points between targets with and without expert support (Appendix~\ref{app:conditioned_correction}).
Format and omission failures also matter: Kimi-Audio has only 35.92\% parseable coaching outputs.

\begin{table}[t]
\centering
\caption{
Open-ended coaching results (\%). Fmt. measures parseable normalized records over 515 recordings, not necessarily non-empty claims or raw JSON compliance.
S. Hit, Hit, and Contra. use all 6,162 diagnosis claims. Valid, Weak, Invalid, and E2E use 2,575 fixed prescription slots.
E2E credits valid advice linked to an expert-supported diagnosis. Invalid excludes vocally inappropriate advice; these omitted labels and missing slots mean Valid, Weak, and Invalid need not sum to 100\%.
Bold marks the group maximum for S. Hit, Hit, Valid, and E2E.}
\label{tab:open_ended_coaching_results}
\scriptsize
\setlength{\tabcolsep}{2.6pt}
\renewcommand{\arraystretch}{1.14}
\begin{adjustbox}{max width=\textwidth}
\begin{tabular}{@{}llcccccccc@{}}
\toprule
\multirow{2}{*}{\textbf{Protocol}} & \multirow{2}{*}{\textbf{Model}} & \multirow{2}{*}{\textbf{Fmt.}} &
\multicolumn{3}{c}{\textbf{Diagnosis}} & \multicolumn{4}{c}{\textbf{Correction}} \\
\cmidrule(lr){4-6}\cmidrule(l){7-10}
& & & \textbf{S. Hit} & \textbf{Hit} & \textbf{Contra. $\downarrow$} & \textbf{Valid} & \textbf{Weak} & \textbf{Invalid $\downarrow$} & \textbf{E2E} \\
\midrule
\multirow{6}{*}{\textsc{Main}} & R1-AQA~\citep{r1aqa} & 100.00 & 0.89 & 14.49 & 0.28 & 24.50 & 59.03 & 16.47 & 8.62 \\
 & Qwen2.5-Omni~\citep{qwen25omni} & 100.00 & 3.10 & \textbf{63.11} & 0.37 & 29.48 & 70.06 & 0.39 & 19.42 \\
 & Kimi-Audio~\citep{kimi_audio} & 35.92 & 1.35 & 20.42 & 0.11 & 19.61 & 15.57 & 0.66 & 12.39 \\
 & MiMo-Audio~\citep{mimo_audio} & 99.61 & 2.99 & 39.21 & 0.41 & 84.70 & 14.83 & 0.04 & 42.68 \\
 & Qwen3-Omni~\citep{qwen3omni} & 100.00 & \textbf{5.86} & 47.94 & 3.78 & \textbf{88.39} & 8.04 & 3.46 & \textbf{53.67} \\
 & Fun-Audio-Chat~\citep{fun_audio_llm} & 96.50 & 5.37 & 57.66 & 0.45 & 66.68 & 28.74 & 1.01 & 47.07 \\
\midrule
\multirow{4}{*}{\textsc{Closed}} & GPT-Audio 1.5~\citep{gpt_audio} & 100.00 & 4.33 & 47.39 & 0.37 & 92.58 & 7.42 & 0.00 & 57.55 \\
 & Gemini 3 Flash Preview~\citep{gemini25flash} & 99.61 & 1.64 & 30.14 & 0.57 & 78.29 & 20.12 & 0.08 & 33.13 \\
 & Qwen3.5-Omni Flash~\citep{qwen35omni} & 99.42 & 6.64 & 48.41 & 0.70 & 94.72 & 4.47 & 0.00 & 58.02 \\
 & Qwen3.5-Omni Plus~\citep{qwen35omni} & 99.22 & \textbf{7.53} & \textbf{52.48} & 0.47 & \textbf{96.85} & 2.29 & 0.00 & \textbf{63.96} \\
\midrule
\multirow{2}{*}{\textsc{Simple}} & Audio Flamingo 3~\citep{audio_flamingo3} & 100.00 & \textbf{1.61} & \textbf{24.63} & 0.16 & \textbf{5.44} & 18.06 & 26.02 & \textbf{2.80} \\
 & Music Flamingo~\citep{music_flamingo} & 100.00 & 0.26 & 2.74 & 0.19 & 4.58 & 2.99 & 12.39 & 0.89 \\
\bottomrule
\end{tabular}
\end{adjustbox}
\end{table}

\section{Conclusion}

We introduced \textbf{VocalCoachBench}, combining controlled same-song comparison, diverse-song segment annotations, and claim-based evaluation of diagnosis and correction.
Expert disagreement motivates evaluating multiple references and levels of specificity.
The strongest models can compare performances and cover broad issue domains, but remain limited in fine-grained issue identification and precise diagnosis.
Separating prescription validity from diagnosis--prescription alignment further reveals that plausible advice need not address an expert-supported problem.
VocalCoachBench provides a public testbed for moving audio-language evaluation beyond description toward grounded expert feedback.

\paragraph{Limitations and future work}
VocalCoachBench covers English solo singing evaluated by professional vocal trainers who wrote their feedback in Korean, their strongest professional language; it does not establish multilingual or cross-cultural generality, and its same-song subset is limited to a single song.
The benchmark is audio-only and one-shot, whereas real coaching may use visual or bodily cues and adapt over repeated attempts, and its segment task assumes expert-provided segments rather than testing end-to-end discovery.
Claim-based scoring measures coverage of expert claims rather than full pedagogical value and may under-credit valid feedback outside them.
Prior exposure to the public diverse-song recordings cannot be ruled out, although removing the 32 recordings sourced from MedleyDB changes each model's Top-3 macro-F1 by at most 1.34 points (Appendix~\ref{app:exposure}).
Because incorrect vocal guidance may encourage strain or unsafe self-training, model outputs should not substitute for professional instruction.
Future work should extend VocalCoachBench toward multilingual, multimodal, interactive, and longitudinal coaching evaluation.

\begin{ack}
This research was supported by Culture, Sports and Tourism R\&D Program through the Korea Creative Content Agency (KOCCA) grant funded by the Ministry of Culture, Sports and Tourism (MCST) in 2026 (Project Name: Development of AI Agent Technology Based on Artists Unique Characteristics for Interactive Cultural Creation, Project Number: RS-2025-02312732, Contribution Rate: 90\%), and by the Artificial intelligence industrial convergence cluster development project funded by the Ministry of Science and ICT (MSIT, Korea) \& Gwangju Metropolitan City.
\end{ack}

\clearpage

%%%%%%%%%%%%%%%%%%%%%%%%%%%%%%%%%%%%%%%%%%%%%%%%%%%%%%%%%%%%
\bibliographystyle{plainnat}
\bibliography{references}

\clearpage

\clearpage

\raggedbottom
\appendix

\section*{Appendix for VocalCoachBench}

\addcontentsline{toc}{section}{Appendix}

\section{Annotation examples}
\label{app:annotation_examples}

This section shows minimal examples of the two annotation products used in VocalCoachBench.
Audio-level coaching reviews are post-processed into atomic claims for open-ended coaching evaluation.
Segment-grounded feedback, used for the structured expert-localized segment issue classification task, is kept as localized feedback with time spans and issue tags; it is not used for claim-based evaluation.

\subsection{Annotation schema}
\label{app:annotation_schema}

\begin{center}
\small
\begin{tabularx}{\linewidth}{p{0.26\linewidth}>{\raggedright\arraybackslash}X}
\toprule
Annotation type & Fields used in evaluation \\
\midrule
Audio-level coaching review & Subset, feedback block, issue tags, and atomic claims.
These claims are used for open-ended diagnosis and correction evaluation. \\
Segment-grounded feedback & Subset, time span, localized feedback, and issue tags.
These annotations support expert-localized segment issue classification and temporal agreement analysis. \\
\bottomrule
\end{tabularx}
\captionof{table}{Two annotation products used in VocalCoachBench and their roles in evaluation.}
\label{tab:annotation_products}
\end{center}

\subsection{Audio-level feedback and claim decomposition examples}
\label{app:audio_level_claim_example}

\begin{center}
\small
\begin{tabularx}{\linewidth}{p{0.18\linewidth}>{\raggedright\arraybackslash}X}
\toprule
Field & Content \\
\midrule
Subset & Same-song \\
Annotation unit & Audio-level coaching feedback block \\
Fine tags & Breath, Vocalization \\
Parent tags & Technical Production \\
Translated feedback & The vocal folds are quite spread apart, so a lot of breath is leaking out and the voice lacks power.
After inhaling, don't make sound right away; practice closing the vocal folds first, then gradually bringing them into contact, feeling the vibration, and producing a straight tone.
The nasality is also strong, so the voice sounds muffled.
Practice vocalization by lifting the soft palate more, relaxing the nose, and making sound while exhaling through the mouth. \\
\bottomrule
\end{tabularx}

\vspace{0.75em}

\begin{tabularx}{\linewidth}{p{0.08\linewidth}p{0.14\linewidth}>{\raggedright\arraybackslash}X p{0.10\linewidth}}
\toprule
Claim ID & Type & Atomic claim & Target \\
\midrule
i1 & diagnosis & The vocal folds are quite spread apart, so a lot of breath is leaking out and the voice lacks power. & -- \\
c1 & correction & After inhaling, do not make sound right away; practice closing the vocal folds first, then gradually bringing them into contact, feeling the vibration, and producing a straight tone. & i1 \\
i2 & diagnosis & The nasality is strong, so the voice sounds muffled. & -- \\
c2 & correction & Practice vocalization by lifting the soft palate more, relaxing the nose, and making sound while exhaling through the mouth. & i2 \\
\bottomrule
\end{tabularx}
\captionof{table}{Example of audio-level claim decomposition.
A translated free-form coaching feedback block is decomposed into atomic claims for open-ended evaluation.}
\label{tab:audio_level_claim_decomposition}
\end{center}

\subsection{Segment-grounded feedback examples}
\label{app:segment_feedback_examples}

\begin{center}
\begin{minipage}{\linewidth}
\small
\begin{tabularx}{\linewidth}{p{0.08\linewidth}p{0.16\linewidth}p{0.18\linewidth}>{\raggedright\arraybackslash}X}
\toprule
ID & Time span & Tags & Translated segment-grounded feedback \\
\midrule
SEG-1 & 0:09.596--0:11.597 & pitch & Do not sing as if the pitch is dropping off abruptly; sing so that it connects and changes naturally. \\
SEG-2 & 2:21.958--2:24.554 & breath, diction & Pronouncing it as ``po hyu'' causes breath to leak, making the pitch unstable and leaving insufficient breath to sustain it.
Pronounce it as ``po yu'' and pay more attention to the vowel. \\
\bottomrule
\end{tabularx}
\captionof{table}{Examples of segment-grounded feedback from the diverse-song subset.
}
\label{tab:segment_feedback_examples}
\end{minipage}
\end{center}

\subsection{Multi-annotator segment overlap}
\label{app:multi_annotator_overlap}

\begin{center}
\small
\begin{tabularx}{\linewidth}{p{0.12\linewidth}p{0.15\linewidth}p{0.18\linewidth}>{\raggedright\arraybackslash}X}
\toprule
Annotator & Span & Tags & Translated segment-grounded feedback \\
\midrule
A & 0:53.310--1:00.849 & breath, technique, pitch & The singer has the technique to sing softly even at high pitches, but appears to waver due to unstable breath.
More consistent breath practice is needed. \\
B & 0:53.477--1:00.777 & breath, pitch & Breath support falls short; as the singer approaches the ending notes, chest voice involvement increases and the overall pitch becomes unstable.
Before that section, the singer should take in enough breath and avoid letting all the breath out from the first word. \\
\bottomrule
\end{tabularx}
\captionof{table}{Example of overlapping segment annotations.
The two annotators selected nearly the same time span (IoU = 0.968) and both identified breath-related instability, while differing in secondary tags and explanation detail.}
\label{tab:multi_annotator_overlap_example}
\end{center}

\section{Score-derived triplet ranking diagnostic}
\label{app:score_derived_triplet_ranking}

The main triplet-ranking evaluation in Table~\ref{tab:structured_protocol_results} uses direct pairwise audio comparison: each triplet is decomposed into three pairwise preference decisions and then evaluated against the expert relative ranking.
We use direct pairwise comparison as the primary protocol because the
gold target is itself a relative judgment, not an absolute quality score.

We additionally evaluate a scalar score-derived protocol as a diagnostic for models or inference paths that do not reliably support multiple audio inputs in a single call.
In this protocol, each recording in a triplet is presented to the
model independently and the model is asked to output an overall singing-quality score on the 0--5 scale.
The three independently elicited scores are then sorted
to induce a triplet ranking.

We report four diagnostics.
Pairwise accuracy (\textbf{Pair}) measures the fraction of the three within-triplet pairwise relations that agree with the expert ordering;
tied pairs and pairs with a missing score are treated as non-decisions and receive no pairwise credit, because repeated scalar scores do not express a discriminative ordering.
Pairwise concordance ($\tau_{\rm pair}$) is the mean of $+1$ for agreement, $-1$ for disagreement, and $0$ for a tied or missing pair; it equals Kendall's $\tau$ for complete, untied rankings.
Exact accuracy (\textbf{Exact}) requires the entire three-item permutation to match the expert ranking.
Tie rate (\textbf{Tie}) is the fraction of within-triplet pairs whose two recordings receive the same score.

\begin{table}[H]
\centering
\caption{
Auxiliary score-derived triplet ranking. Values are obtained by independently eliciting scalar quality scores for the three recordings in each triplet and sorting the recordings by those scores.
Tie\,($\downarrow$) is the fraction of within-triplet pairs with tied scores.
All entries except $\tau_{\rm pair}$ are percentages. Bold and underline mark the best and second-best values within each protocol group.
}
\label{tab:score_derived_triplet_ranking}
\footnotesize
\setlength{\tabcolsep}{4.0pt}
\renewcommand{\arraystretch}{1.12}
\begin{adjustbox}{max width=\textwidth}
\begin{tabular}{@{}llcccc@{}}
\toprule
\textbf{Protocol} & \textbf{Model} & \textbf{Pair} & \textbf{$\tau_{\rm pair}$} & \textbf{Exact} & \textbf{Tie $\downarrow$} \\
\midrule
\rowcolor{gray!10}
\textsc{Baseline} & Random permutation & 50.0 & .000 & 16.7 & 0.0 \\
\midrule
\multirow{6}{*}{\textsc{Main}} & R1-AQA~\citep{r1aqa} & 0.0 & .000 & 0.0 & 100.0 \\
& Qwen2.5-Omni~\citep{qwen25omni} & 4.4 & .014 & 0.0 & 92.6 \\
& Kimi-Audio~\citep{kimi_audio} & \underline{33.5} & .037 & \textbf{6.1} & \underline{30.7} \\
& MiMo-Audio~\citep{mimo_audio} & \textbf{36.9} & \textbf{.182} & 1.1 & 44.3 \\
& Qwen3-Omni~\citep{qwen3omni} & 13.6 & \underline{.074} & 0.3 & 80.2 \\
& Fun-Audio-Chat~\citep{fun_audio_llm} & 31.6 & $-$.004 & \underline{5.8} & \textbf{27.3} \\
\midrule
\multirow{4}{*}{\textsc{Closed}} & GPT-Audio 1.5~\citep{gpt_audio} & \textbf{43.7} & \textbf{.302} & \underline{6.6} & \underline{42.7} \\
& Gemini 3 Flash Preview~\citep{gemini25flash} & 37.7 & .157 & \textbf{9.5} & \textbf{40.4} \\
& Qwen3.5-Omni Flash~\citep{qwen35omni} & 28.1 & .152 & 1.6 & 58.9 \\
& Qwen3.5-Omni Plus~\citep{qwen35omni} & \underline{40.9} & \underline{.265} & 5.0 & 44.6 \\
\midrule
\multirow{2}{*}{\textsc{Simple}} & Audio Flamingo 3~\citep{audio_flamingo3} & \underline{0.0} & \underline{.000} & 0.0 & \underline{100.0} \\
& Music Flamingo~\citep{music_flamingo} & \textbf{28.4} & \textbf{.233} & 0.0 & \textbf{66.5} \\
\bottomrule
\end{tabular}
\end{adjustbox}
\end{table}

The score-derived protocol is substantially weaker than direct pairwise
comparison.
Several models collapse to repeated scores on a large fraction of within-triplet pairs: R1-AQA and Audio Flamingo 3 tie on every pair, Qwen2.5-Omni ties on 92.6\%, Qwen3-Omni on 80.2\%, and Music Flamingo on 66.5\%.
This score collapse prevents the induced ranking from representing a meaningful comparison even when the model may be capable of making relative judgments.
Closed-source models produce fewer ties, but their pairwise accuracy remains below the random baseline of 50\%, with the best score-derived result reaching 43.7\% pairwise accuracy and $\tau_{\rm pair}=.302$.
These results support treating scalar score elicitation as an auxiliary diagnostic rather than the main triplet ranking protocol.

\section{Category-level and subset results}
\label{app:category_analysis}

This section reports class-level and subset results for the structured tasks in Table~\ref{tab:structured_protocol_results}, computed from the same stored responses, parsers, and recording-balanced expert references.
Per-class precision, recall, and F1 treat each category as a one-vs-rest decision; their unweighted mean is the macro-F1 in the main results.
Prediction frequency is the percentage of recordings whose predicted Top-3 set contains a category, and first-label confusion compares each model's first-ranked label with each expert's first-ranked label, averaging references within each recording.
Differences from the majority baseline use paired bootstrap intervals and are not adjusted for multiple comparisons.

No model exceeds the majority baseline significantly in Top-3 macro-F1, whereas five models do so in segment macro-F1 (Table~\ref{tab:majority_bootstrap_subsets}).
Averaged over the 12 models, F1@3 is similar on the same-song and diverse-song subsets (39.78\% and 39.31\%), whereas macro-F1 is higher on diverse-song recordings (29.56\% vs.\ 27.18\%).
The majority baseline is 15.33 F1@3 points higher on same-song recordings, reflecting their more concentrated vocalization, breath support, and pitch labels (Figure~\ref{fig:dataset_analysis}a); model differences vary in direction, for example 12.12 points for Gemini 3 Flash Preview and $-$5.31 points for R1-AQA.

Prediction frequencies depart from reference prevalence in both directions (Table~\ref{tab:category_freq}).
Rhythm is the least prevalent reference category (18.9\%), yet six models include it for more than 90\% of recordings; six models never predict expression, and nine predict it for at most 1.6\% of recordings.
Several models follow near-fixed label sets: R1-AQA predicts pitch, rhythm, and diction for every recording, and Qwen3.5-Omni Plus predicts pitch, rhythm, and technique for more than 96\% of recordings.
The highest per-class F1 in six of the seven categories is obtained by a model that predicts that category for at least 96\% of recordings (Table~\ref{tab:category_f1}).
First-label confusions show the same pattern (Tables~\ref{tab:top1_confusion} and~\ref{tab:top1_confusion_full}): for expert first-ranked vocalization, the most frequent model first label is pitch for eight models and rhythm for three, and Fun-Audio-Chat assigns vocalization to most recordings whose expert first label is breath support.
In segment classification, no model produces a true positive for technique or expression (Table~\ref{tab:segment_category_f1}).
Expert--expert category agreement is highest for vocalization (81.76\%) and lowest for rhythm (26.78\%) and technique (25.63\%).

\begin{table}[H]
\centering
\caption{Comparison with the majority baseline and subset results (\%). $\Delta$ is the model minus the majority baseline in macro-F1 points, with paired percentile bootstrap 95\% intervals over 10,000 recording resamples; segment events are resampled with their recordings. Same and Diverse use the 207 same-song and 308 diverse-song recordings.}
\label{tab:majority_bootstrap_subsets}
\scriptsize
\setlength{\tabcolsep}{3pt}
\renewcommand{\arraystretch}{1.08}
\begin{adjustbox}{max width=\textwidth}
\begin{tabular}{@{}llcccccc@{}}
\toprule
\textbf{Protocol} & \textbf{Model} & \textbf{\shortstack{Top-3 macro-F1\\$\Delta$ [95\% CI]}} & \textbf{\shortstack{Segment macro-F1\\$\Delta$ [95\% CI]}} & \textbf{\shortstack{F1@3\\Same}} & \textbf{\shortstack{F1@3\\Diverse}} & \textbf{\shortstack{Macro-F1\\Same}} & \textbf{\shortstack{Macro-F1\\Diverse}} \\
\midrule
\multicolumn{2}{@{}l}{Majority (no audio)} & -- & -- & 71.66 & 56.33 & 35.53 & 30.62 \\
\midrule
\multirow{6}{*}{\textsc{Main}} & R1-AQA & $-$11.28 [$-$12.26, $-$10.32] & $-$1.23 [$-$2.77, 0.34] & 31.32 & 36.63 & 18.92 & 22.71 \\
 & Qwen2.5-Omni & $-$1.65 [$-$3.13, $-$0.15] & $-$0.56 [$-$2.49, 1.68] & 40.93 & 36.58 & 31.26 & 29.32 \\
 & Kimi-Audio & $-$4.47 [$-$6.12, $-$2.84] & +5.40 [2.37, 8.23] & 32.61 & 34.63 & 26.41 & 28.25 \\
 & MiMo-Audio & $-$2.57 [$-$4.01, $-$1.13] & +0.00 [0.00, 0.00] & 41.12 & 43.51 & 26.96 & 31.86 \\
 & Qwen3-Omni & $-$0.39 [$-$1.88, 1.08] & +5.97 [2.42, 9.68] & 38.19 & 42.69 & 29.03 & 33.53 \\
 & Fun-Audio-Chat & $-$3.01 [$-$4.60, $-$1.44] & +0.88 [$-$1.03, 2.83] & 40.15 & 42.21 & 26.84 & 31.11 \\
\midrule
\multirow{4}{*}{\textsc{Closed}} & GPT-Audio 1.5 & +0.74 [$-$0.91, 2.30] & $-$0.14 [$-$2.59, 2.68] & 40.31 & 38.31 & 30.27 & 33.21 \\
 & Gemini 3 Flash Preview & +0.68 [$-$0.32, 1.71] & +4.55 [1.46, 7.97] & 57.09 & 44.97 & 35.32 & 31.52 \\
 & Qwen3.5-Omni Flash & $-$1.10 [$-$2.50, 0.23] & +6.42 [2.51, 9.99] & 47.69 & 42.42 & 31.71 & 31.40 \\
 & Qwen3.5-Omni Plus & $-$11.19 [$-$12.49, $-$9.87] & +0.52 [$-$2.11, 3.46] & 32.96 & 31.17 & 19.88 & 21.93 \\
\midrule
\multirow{2}{*}{\textsc{Simple}} & Audio Flamingo 3 & $-$9.44 [$-$10.71, $-$8.22] & +5.30 [2.70, 8.03] & 31.03 & 35.44 & 20.54 & 24.77 \\
 & Music Flamingo & +0.60 [$-$1.16, 2.35] & $-$4.41 [$-$5.85, $-$2.90] & 43.96 & 43.13 & 29.09 & 35.04 \\
\midrule
\multicolumn{2}{@{}l}{Mean of 12 models} & -- & -- & 39.78 & 39.31 & 27.18 & 29.56 \\
\bottomrule
\end{tabular}
\end{adjustbox}
\end{table}

\begin{table}[H]
\centering
\caption{Top-3 per-class F1 (\%) over 515 recordings. Macro is the unweighted mean of the seven classes. The expert--expert row compares distinct ordered expert pairs within the 513 recordings with at least two reviews.}
\label{tab:category_f1}
\scriptsize
\setlength{\tabcolsep}{3pt}
\renewcommand{\arraystretch}{1.08}
\begin{adjustbox}{max width=\textwidth}
\begin{tabular}{@{}llcccccccc@{}}
\toprule
\textbf{Protocol} & \textbf{Model} & \textbf{Breath} & \textbf{Vocal.} & \textbf{Tech.} & \textbf{Pitch} & \textbf{Rhythm} & \textbf{Diction} & \textbf{Expr.} & \textbf{Macro} \\
\midrule
\multicolumn{2}{@{}l}{Majority (no audio)} & 76.6 & 86.5 & 0.0 & 65.9 & 0.0 & 0.0 & 0.0 & 32.7 \\
\midrule
\multirow{6}{*}{\textsc{Main}} & R1-AQA & 0.0 & 0.0 & 0.0 & 65.9 & 31.8 & 52.4 & 0.0 & 21.4 \\
 & Qwen2.5-Omni & 17.3 & 22.7 & 35.7 & 65.9 & 28.5 & 47.4 & 0.0 & 31.1 \\
 & Kimi-Audio & 15.9 & 44.0 & 32.4 & 21.1 & 31.6 & 52.1 & 0.6 & 28.2 \\
 & MiMo-Audio & 0.6 & 82.3 & 37.5 & 14.0 & 27.4 & 49.1 & 0.0 & 30.1 \\
 & Qwen3-Omni & 0.0 & 69.1 & 30.9 & 42.3 & 31.6 & 52.3 & 0.0 & 32.3 \\
 & Fun-Audio-Chat & 3.9 & 85.1 & 30.7 & 12.9 & 31.4 & 39.3 & 4.6 & 29.7 \\
\midrule
\multirow{4}{*}{\textsc{Closed}} & GPT-Audio 1.5 & 43.5 & 0.5 & 41.5 & 65.9 & 17.7 & 28.8 & 36.3 & 33.5 \\
 & Gemini 3 Flash Preview & 48.8 & 71.2 & 42.6 & 65.3 & 1.0 & 1.1 & 3.8 & 33.4 \\
 & Qwen3.5-Omni Flash & 73.6 & 18.7 & 11.1 & 65.1 & 29.0 & 21.5 & 2.4 & 31.6 \\
 & Qwen3.5-Omni Plus & 0.0 & 1.5 & 41.9 & 65.8 & 31.4 & 10.1 & 0.0 & 21.5 \\
\midrule
\multirow{2}{*}{\textsc{Simple}} & Audio Flamingo 3 & 0.0 & 1.5 & 13.1 & 65.5 & 31.8 & 51.1 & 0.0 & 23.3 \\
 & Music Flamingo & 75.6 & 5.2 & 20.0 & 60.5 & 30.8 & 24.8 & 16.4 & 33.3 \\
\midrule
\multicolumn{2}{@{}l}{Expert--expert} & 65.5 & 81.8 & 25.6 & 63.4 & 26.8 & 37.7 & 39.6 & 48.6 \\
\bottomrule
\end{tabular}
\end{adjustbox}
\end{table}

\begin{table}[H]
\centering
\caption{Top-3 per-class precision / recall (\%) over 515 recordings.}
\label{tab:category_pr}
\scriptsize
\setlength{\tabcolsep}{2pt}
\renewcommand{\arraystretch}{1.08}
\begin{adjustbox}{max width=\textwidth}
\begin{tabular}{@{}llccccccc@{}}
\toprule
\textbf{Protocol} & \textbf{Model} & \textbf{Breath} & \textbf{Vocal.} & \textbf{Tech.} & \textbf{Pitch} & \textbf{Rhythm} & \textbf{Diction} & \textbf{Expr.} \\
\midrule
\multicolumn{2}{@{}l}{Majority (no audio)} & 62.1\,/\,100.0 & 76.2\,/\,100.0 & 0.0\,/\,0.0 & 49.1\,/\,100.0 & 0.0\,/\,0.0 & 0.0\,/\,0.0 & 0.0\,/\,0.0 \\
\midrule
\multirow{6}{*}{\textsc{Main}} & R1-AQA & 0.0\,/\,0.0 & 0.0\,/\,0.0 & 0.0\,/\,0.0 & 49.1\,/\,100.0 & 18.9\,/\,100.0 & 35.5\,/\,100.0 & 0.0\,/\,0.0 \\
 & Qwen2.5-Omni & 74.6\,/\,9.8 & 77.9\,/\,13.3 & 27.8\,/\,49.7 & 49.1\,/\,100.0 & 18.6\,/\,60.8 & 35.9\,/\,69.8 & 0.0\,/\,0.0 \\
 & Kimi-Audio & 65.9\,/\,9.1 & 81.9\,/\,30.1 & 24.7\,/\,47.3 & 53.5\,/\,13.1 & 18.8\,/\,98.6 & 35.4\,/\,98.6 & 50.0\,/\,0.3 \\
 & MiMo-Audio & 100.0\,/\,0.3 & 76.0\,/\,89.8 & 29.0\,/\,53.2 & 52.6\,/\,8.1 & 17.2\,/\,68.2 & 35.8\,/\,78.3 & 0.0\,/\,0.0 \\
 & Qwen3-Omni & 0.0\,/\,0.0 & 75.2\,/\,64.0 & 33.9\,/\,28.3 & 50.7\,/\,36.3 & 18.8\,/\,99.0 & 38.1\,/\,83.8 & 0.0\,/\,0.0 \\
 & Fun-Audio-Chat & 50.0\,/\,2.0 & 76.3\,/\,96.2 & 26.0\,/\,37.6 & 45.2\,/\,7.5 & 18.8\,/\,94.9 & 33.3\,/\,48.0 & 28.6\,/\,2.5 \\
\midrule
\multirow{4}{*}{\textsc{Closed}} & GPT-Audio 1.5 & 65.7\,/\,32.5 & 100.0\,/\,0.3 & 26.7\,/\,92.8 & 49.1\,/\,100.0 & 14.8\,/\,22.1 & 41.8\,/\,22.0 & 37.8\,/\,34.9 \\
 & Gemini 3 Flash Preview & 62.8\,/\,39.9 & 79.6\,/\,64.5 & 27.1\,/\,100.0 & 49.2\,/\,97.2 & 50.0\,/\,0.5 & 50.0\,/\,0.5 & 39.6\,/\,2.0 \\
 & Qwen3.5-Omni Flash & 61.3\,/\,92.2 & 75.0\,/\,10.7 & 23.8\,/\,7.2 & 49.4\,/\,95.5 & 18.1\,/\,72.8 & 39.1\,/\,14.8 & 28.6\,/\,1.2 \\
 & Qwen3.5-Omni Plus & 0.0\,/\,0.0 & 100.0\,/\,0.8 & 26.8\,/\,96.0 & 49.1\,/\,99.4 & 18.7\,/\,97.4 & 62.5\,/\,5.5 & 0.0\,/\,0.0 \\
\midrule
\multirow{2}{*}{\textsc{Simple}} & Audio Flamingo 3 & 0.0\,/\,0.0 & 100.0\,/\,0.8 & 25.9\,/\,8.8 & 49.6\,/\,96.4 & 18.9\,/\,99.0 & 35.6\,/\,90.1 & 0.0\,/\,0.0 \\
 & Music Flamingo & 62.0\,/\,96.9 & 70.0\,/\,2.7 & 26.6\,/\,16.1 & 48.9\,/\,79.4 & 19.0\,/\,80.3 & 37.8\,/\,18.4 & 45.7\,/\,10.0 \\
\bottomrule
\end{tabular}
\end{adjustbox}
\end{table}

\begin{table}[H]
\centering
\caption{Top-3 prediction frequency (\%): the percentage of the 515 recordings whose predicted Top-3 set contains each category. Reference prevalence is the recording-balanced share of expert Top-3 sets containing the category. Missing or unparsable outputs contribute no label.}
\label{tab:category_freq}
\scriptsize
\setlength{\tabcolsep}{3pt}
\renewcommand{\arraystretch}{1.08}
\begin{adjustbox}{max width=\textwidth}
\begin{tabular}{@{}llccccccc@{}}
\toprule
\textbf{Protocol} & \textbf{Model} & \textbf{Breath} & \textbf{Vocal.} & \textbf{Tech.} & \textbf{Pitch} & \textbf{Rhythm} & \textbf{Diction} & \textbf{Expr.} \\
\midrule
\multicolumn{2}{@{}l}{Reference prevalence} & 62.1 & 76.2 & 27.0 & 49.1 & 18.9 & 35.5 & 31.2 \\
\multicolumn{2}{@{}l}{Majority (no audio)} & 100.0 & 100.0 & 0.0 & 100.0 & 0.0 & 0.0 & 0.0 \\
\midrule
\multirow{6}{*}{\textsc{Main}} & R1-AQA & 0.0 & 0.0 & 0.0 & 100.0 & 100.0 & 100.0 & 0.0 \\
 & Qwen2.5-Omni & 8.2 & 13.0 & 48.2 & 100.0 & 61.7 & 68.9 & 0.0 \\
 & Kimi-Audio & 8.5 & 28.0 & 51.7 & 12.0 & 99.0 & 98.8 & 0.2 \\
 & MiMo-Audio & 0.2 & 90.1 & 49.5 & 7.6 & 75.0 & 77.7 & 0.0 \\
 & Qwen3-Omni & 0.0 & 64.9 & 22.5 & 35.1 & 99.4 & 78.1 & 0.0 \\
 & Fun-Audio-Chat & 2.5 & 96.1 & 39.0 & 8.2 & 95.1 & 51.1 & 2.7 \\
\midrule
\multirow{4}{*}{\textsc{Closed}} & GPT-Audio 1.5 & 30.7 & 0.2 & 93.6 & 100.0 & 28.2 & 18.6 & 28.7 \\
 & Gemini 3 Flash Preview & 39.4 & 61.7 & 99.6 & 97.1 & 0.2 & 0.4 & 1.6 \\
 & Qwen3.5-Omni Flash & 93.4 & 10.9 & 8.2 & 95.0 & 76.1 & 13.4 & 1.4 \\
 & Qwen3.5-Omni Plus & 0.4 & 0.6 & 96.5 & 99.4 & 98.3 & 3.1 & 0.0 \\
\midrule
\multirow{2}{*}{\textsc{Simple}} & Audio Flamingo 3 & 0.2 & 0.6 & 9.1 & 95.5 & 98.8 & 89.7 & 0.0 \\
 & Music Flamingo & 97.1 & 2.9 & 16.3 & 79.8 & 79.8 & 17.3 & 6.8 \\
\bottomrule
\end{tabular}
\end{adjustbox}
\end{table}

\begin{table}[H]
\centering
\caption{First-label confusion. For each expert first-ranked label (columns), the cell gives the most frequent model first-ranked label and the percentage of that expert label it receives, averaging expert references within each recording. Codes: B breath support, V vocalization, T technique, P pitch, R rhythm, D diction, E expression; -- missing output.}
\label{tab:top1_confusion}
\scriptsize
\setlength{\tabcolsep}{2.5pt}
\renewcommand{\arraystretch}{1.08}
\begin{adjustbox}{max width=\textwidth}
\begin{tabular}{@{}llccccccc@{}}
\toprule
\textbf{Protocol} & \textbf{Model} & \textbf{\shortstack{Expert\\Breath}} & \textbf{\shortstack{Expert\\Vocal.}} & \textbf{\shortstack{Expert\\Tech.}} & \textbf{\shortstack{Expert\\Pitch}} & \textbf{\shortstack{Expert\\Rhythm}} & \textbf{\shortstack{Expert\\Diction}} & \textbf{\shortstack{Expert\\Expr.}} \\
\midrule
\multirow{6}{*}{\textsc{Main}} & R1-AQA & P\,100 & P\,100 & P\,100 & P\,100 & P\,100 & P\,100 & P\,100 \\
 & Qwen2.5-Omni & P\,61 & P\,68 & P\,65 & P\,68 & P\,59 & P\,64 & P\,50 \\
 & Kimi-Audio & R\,75 & R\,75 & R\,79 & R\,66 & R\,67 & R\,80 & R\,65 \\
 & MiMo-Audio & R\,55 & R\,55 & R\,47 & R\,62 & R\,46 & R\,51 & V\,49 \\
 & Qwen3-Omni & R\,41 & R\,38 & R\,36 & R\,38 & P\,38 & R\,42 & R\,38 \\
 & Fun-Audio-Chat & V\,96 & V\,95 & V\,95 & V\,93 & V\,87 & V\,86 & V\,91 \\
\midrule
\multirow{4}{*}{\textsc{Closed}} & GPT-Audio 1.5 & P\,100 & P\,99 & P\,100 & P\,100 & P\,100 & P\,99 & P\,100 \\
 & Gemini 3 Flash Preview & P\,89 & P\,91 & P\,91 & P\,94 & P\,97 & P\,97 & P\,94 \\
 & Qwen3.5-Omni Flash & P\,89 & P\,90 & P\,86 & P\,90 & P\,89 & P\,81 & P\,88 \\
 & Qwen3.5-Omni Plus & P\,100 & P\,99 & P\,98 & P\,99 & P\,100 & P\,94 & P\,98 \\
\midrule
\multirow{2}{*}{\textsc{Simple}} & Audio Flamingo 3 & P\,95 & P\,94 & P\,95 & P\,99 & P\,100 & P\,93 & P\,96 \\
 & Music Flamingo & P\,80 & P\,77 & P\,85 & P\,81 & P\,87 & P\,83 & P\,78 \\
\bottomrule
\end{tabular}
\end{adjustbox}
\end{table}

\begin{table}[H]
\centering
\caption{Full first-label confusion matrices (\%). Rows are expert first-ranked labels and columns are model first-ranked labels; each row is normalized within the expert label, averaging expert references within each recording, so rows sum to 100 up to rounding. Codes as in Table~\ref{tab:top1_confusion}; -- missing output. Recording-balanced expert first-label masses over the 515 recordings: B 121.3, V 179.3, T 32.0, P 71.8, R 18.7, D 51.8, E 40.0.}
\label{tab:top1_confusion_full}
\scriptsize
\setlength{\tabcolsep}{2.5pt}
\renewcommand{\arraystretch}{1.0}
\begin{adjustbox}{max width=\textwidth}
\begin{tabular}{@{}l*{8}{c}@{\hspace{10pt}}l*{8}{c}@{}}
\toprule
\multicolumn{9}{@{}l}{\textbf{R1-AQA}} & \multicolumn{9}{l@{}}{\textbf{Qwen2.5-Omni}} \\
Expert & B & V & T & P & R & D & E & -- & Expert & B & V & T & P & R & D & E & -- \\
\cmidrule(r){1-9}\cmidrule(l){10-18}
Breath & 0 & 0 & 0 & 100 & 0 & 0 & 0 & 0 & Breath & 0 & 0 & 0 & 61 & 39 & 0 & 0 & 0 \\
Vocal. & 0 & 0 & 0 & 100 & 0 & 0 & 0 & 0 & Vocal. & 0 & 0 & 0 & 68 & 32 & 0 & 0 & 0 \\
Tech. & 0 & 0 & 0 & 100 & 0 & 0 & 0 & 0 & Tech. & 0 & 0 & 0 & 65 & 35 & 0 & 0 & 0 \\
Pitch & 0 & 0 & 0 & 100 & 0 & 0 & 0 & 0 & Pitch & 0 & 0 & 0 & 68 & 32 & 0 & 0 & 0 \\
Rhythm & 0 & 0 & 0 & 100 & 0 & 0 & 0 & 0 & Rhythm & 0 & 0 & 0 & 59 & 41 & 0 & 0 & 0 \\
Diction & 0 & 0 & 0 & 100 & 0 & 0 & 0 & 0 & Diction & 0 & 0 & 0 & 64 & 36 & 0 & 0 & 0 \\
Expr. & 0 & 0 & 0 & 100 & 0 & 0 & 0 & 0 & Expr. & 0 & 0 & 0 & 50 & 50 & 0 & 0 & 0 \\
\midrule
\multicolumn{9}{@{}l}{\textbf{Kimi-Audio}} & \multicolumn{9}{l@{}}{\textbf{MiMo-Audio}} \\
Expert & B & V & T & P & R & D & E & -- & Expert & B & V & T & P & R & D & E & -- \\
\cmidrule(r){1-9}\cmidrule(l){10-18}
Breath & $<$1 & 0 & 1 & $<$1 & 75 & 23 & 0 & 1 & Breath & 0 & 39 & 0 & 6 & 55 & 0 & 0 & 0 \\
Vocal. & 0 & 0 & 0 & 1 & 75 & 24 & 0 & 1 & Vocal. & 0 & 37 & 0 & 8 & 55 & 0 & 0 & 0 \\
Tech. & 3 & 0 & 0 & 0 & 79 & 18 & 0 & 0 & Tech. & 0 & 44 & 0 & 9 & 47 & 0 & 0 & 0 \\
Pitch & 1 & 0 & 1 & 1 & 66 & 31 & 0 & 0 & Pitch & 0 & 33 & 0 & 5 & 62 & 0 & 0 & 0 \\
Rhythm & 0 & 0 & 3 & 3 & 67 & 28 & 0 & 0 & Rhythm & 0 & 46 & 0 & 8 & 46 & 0 & 0 & 0 \\
Diction & 0 & 0 & 1 & 1 & 80 & 18 & 0 & 0 & Diction & 0 & 38 & 0 & 12 & 51 & 0 & 0 & 0 \\
Expr. & 2 & 0 & 0 & 0 & 65 & 31 & 0 & 1 & Expr. & 0 & 49 & 0 & 10 & 41 & 0 & 0 & 0 \\
\midrule
\multicolumn{9}{@{}l}{\textbf{Qwen3-Omni}} & \multicolumn{9}{l@{}}{\textbf{Fun-Audio-Chat}} \\
Expert & B & V & T & P & R & D & E & -- & Expert & B & V & T & P & R & D & E & -- \\
\cmidrule(r){1-9}\cmidrule(l){10-18}
Breath & 0 & 24 & 1 & 34 & 41 & 0 & 0 & 0 & Breath & 0 & 96 & 0 & 1 & 1 & 0 & 1 & 1 \\
Vocal. & 0 & 25 & 4 & 32 & 38 & $<$1 & 0 & 0 & Vocal. & 0 & 95 & 0 & 0 & 2 & 0 & 2 & 1 \\
Tech. & 0 & 34 & 3 & 27 & 36 & 0 & 0 & 0 & Tech. & 0 & 95 & 0 & 0 & 0 & 0 & 3 & 2 \\
Pitch & 0 & 29 & 2 & 31 & 38 & 0 & 0 & 0 & Pitch & 0 & 93 & 1 & 0 & 3 & 0 & 1 & 1 \\
Rhythm & 0 & 27 & 5 & 38 & 30 & 0 & 0 & 0 & Rhythm & 0 & 87 & 0 & 3 & 5 & 0 & 3 & 3 \\
Diction & 0 & 21 & 0 & 35 & 42 & 2 & 0 & 0 & Diction & 0 & 86 & 0 & 2 & 5 & 0 & 7 & 0 \\
Expr. & 0 & 36 & 2 & 22 & 38 & 1 & 0 & 0 & Expr. & 0 & 91 & 0 & 1 & 2 & 0 & 5 & 0 \\
\midrule
\multicolumn{9}{@{}l}{\textbf{GPT-Audio 1.5}} & \multicolumn{9}{l@{}}{\textbf{Gemini 3 Flash Preview}} \\
Expert & B & V & T & P & R & D & E & -- & Expert & B & V & T & P & R & D & E & -- \\
\cmidrule(r){1-9}\cmidrule(l){10-18}
Breath & 0 & 0 & 0 & 100 & 0 & 0 & 0 & 0 & Breath & 0 & 0 & 11 & 89 & 0 & 0 & 0 & 0 \\
Vocal. & 0 & 0 & 0 & 99 & 1 & $<$1 & 0 & 0 & Vocal. & 0 & $<$1 & 8 & 91 & 0 & 0 & 0 & 0 \\
Tech. & 0 & 0 & 0 & 100 & 0 & 0 & 0 & 0 & Tech. & 0 & 2 & 7 & 91 & 0 & 0 & 0 & 0 \\
Pitch & 0 & 0 & 0 & 100 & 0 & 0 & 0 & 0 & Pitch & 0 & 0 & 6 & 94 & 0 & 0 & 0 & 0 \\
Rhythm & 0 & 0 & 0 & 100 & 0 & 0 & 0 & 0 & Rhythm & 0 & 0 & 3 & 97 & 0 & 0 & 0 & 0 \\
Diction & 0 & 0 & 0 & 99 & 0 & 1 & 0 & 0 & Diction & 0 & 0 & 3 & 97 & 0 & 0 & 0 & 0 \\
Expr. & 0 & 0 & 0 & 100 & 0 & 0 & 0 & 0 & Expr. & 0 & 0 & 6 & 94 & 0 & 0 & 0 & 0 \\
\midrule
\multicolumn{9}{@{}l}{\textbf{Qwen3.5-Omni Flash}} & \multicolumn{9}{l@{}}{\textbf{Qwen3.5-Omni Plus}} \\
Expert & B & V & T & P & R & D & E & -- & Expert & B & V & T & P & R & D & E & -- \\
\cmidrule(r){1-9}\cmidrule(l){10-18}
Breath & 5 & 0 & 0 & 89 & 5 & 1 & 0 & $<$1 & Breath & 0 & 0 & 0 & 100 & 0 & 0 & 0 & $<$1 \\
Vocal. & 6 & 0 & 0 & 90 & 4 & 1 & 0 & 0 & Vocal. & 0 & 0 & 0 & 99 & 1 & 0 & 0 & 0 \\
Tech. & 11 & 0 & 0 & 86 & 3 & 0 & 0 & 0 & Tech. & 0 & 0 & 2 & 98 & 0 & 0 & 0 & 0 \\
Pitch & 6 & 0 & 0 & 90 & 3 & 0 & 0 & 1 & Pitch & 0 & 0 & 0 & 99 & 0 & 0 & 0 & 1 \\
Rhythm & 5 & 0 & 0 & 89 & 5 & 0 & 0 & 0 & Rhythm & 0 & 0 & 0 & 100 & 0 & 0 & 0 & 0 \\
Diction & 11 & 0 & 0 & 81 & 5 & 0 & 0 & 4 & Diction & 0 & 0 & 1 & 94 & 1 & 0 & 0 & 4 \\
Expr. & 8 & 0 & 0 & 88 & 4 & 1 & 0 & 0 & Expr. & 0 & 0 & 2 & 98 & 0 & 0 & 0 & 0 \\
\midrule
\multicolumn{9}{@{}l}{\textbf{Audio Flamingo 3}} & \multicolumn{9}{l@{}}{\textbf{Music Flamingo}} \\
Expert & B & V & T & P & R & D & E & -- & Expert & B & V & T & P & R & D & E & -- \\
\cmidrule(r){1-9}\cmidrule(l){10-18}
Breath & 0 & $<$1 & $<$1 & 95 & 3 & 0 & 0 & 1 & Breath & 0 & 5 & 0 & 80 & 0 & 16 & 0 & 0 \\
Vocal. & 0 & $<$1 & $<$1 & 94 & 5 & 0 & 0 & $<$1 & Vocal. & 0 & 2 & 0 & 77 & 0 & 21 & 0 & 0 \\
Tech. & 0 & 0 & 0 & 95 & 5 & 0 & 0 & 0 & Tech. & 0 & 5 & 0 & 85 & 0 & 10 & 0 & 0 \\
Pitch & 1 & 0 & 0 & 99 & 0 & 0 & 0 & 1 & Pitch & 0 & 3 & 0 & 81 & 0 & 16 & 0 & 0 \\
Rhythm & 0 & 0 & 0 & 100 & 0 & 0 & 0 & 0 & Rhythm & 0 & 3 & 0 & 87 & 0 & 11 & 0 & 0 \\
Diction & 0 & 0 & 2 & 93 & 1 & 0 & 0 & 4 & Diction & 0 & 2 & 0 & 83 & 0 & 15 & 0 & 0 \\
Expr. & 1 & 0 & 0 & 96 & 2 & 0 & 0 & 0 & Expr. & 0 & 4 & 0 & 78 & 0 & 19 & 0 & 0 \\
\bottomrule
\end{tabular}
\end{adjustbox}
\end{table}

\begin{table}[H]
\centering
\caption{Segment per-class F1 (\%) over the 262 consensus events, treating all shared reference labels as positives. Reference prevalence is the percentage of events carrying each label.}
\label{tab:segment_category_f1}
\scriptsize
\setlength{\tabcolsep}{3pt}
\renewcommand{\arraystretch}{1.08}
\begin{adjustbox}{max width=\textwidth}
\begin{tabular}{@{}llcccccccc@{}}
\toprule
\textbf{Protocol} & \textbf{Model} & \textbf{Breath} & \textbf{Vocal.} & \textbf{Tech.} & \textbf{Pitch} & \textbf{Rhythm} & \textbf{Diction} & \textbf{Expr.} & \textbf{Macro} \\
\midrule
\multicolumn{2}{@{}l}{Reference prevalence} & 25.6 & 40.5 & 6.1 & 32.4 & 8.8 & 14.1 & 14.1 & -- \\
\multicolumn{2}{@{}l}{Majority (no audio)} & 0.0 & 57.6 & 0.0 & 0.0 & 0.0 & 0.0 & 0.0 & 8.2 \\
\midrule
\multirow{6}{*}{\textsc{Main}} & R1-AQA & 0.0 & 0.0 & 0.0 & 49.0 & 0.0 & 0.0 & 0.0 & 7.0 \\
 & Qwen2.5-Omni & 0.0 & 1.8 & 0.0 & 46.9 & 0.0 & 5.0 & 0.0 & 7.7 \\
 & Kimi-Audio & 2.7 & 47.3 & 0.0 & 0.0 & 20.0 & 25.4 & 0.0 & 13.6 \\
 & MiMo-Audio & 0.0 & 57.6 & 0.0 & 0.0 & 0.0 & 0.0 & 0.0 & 8.2 \\
 & Qwen3-Omni & 0.0 & 24.8 & 0.0 & 48.8 & 16.7 & 9.1 & 0.0 & 14.2 \\
 & Fun-Audio-Chat & 0.0 & 15.0 & 0.0 & 48.8 & 0.0 & 0.0 & 0.0 & 9.1 \\
\midrule
\multirow{4}{*}{\textsc{Closed}} & GPT-Audio 1.5 & 0.0 & 0.0 & 0.0 & 46.4 & 0.0 & 10.3 & 0.0 & 8.1 \\
 & Gemini 3 Flash Preview & 5.0 & 29.2 & 0.0 & 47.3 & 8.0 & 0.0 & 0.0 & 12.8 \\
 & Qwen3.5-Omni Flash & 22.4 & 3.4 & 0.0 & 48.2 & 28.6 & 0.0 & 0.0 & 14.7 \\
 & Qwen3.5-Omni Plus & 0.0 & 0.0 & 0.0 & 49.1 & 12.1 & 0.0 & 0.0 & 8.7 \\
\midrule
\multirow{2}{*}{\textsc{Simple}} & Audio Flamingo 3 & 0.0 & 35.7 & 0.0 & 49.3 & 0.0 & 9.8 & 0.0 & 13.5 \\
 & Music Flamingo & 0.0 & 1.9 & 0.0 & 0.0 & 0.0 & 24.9 & 0.0 & 3.8 \\
\bottomrule
\end{tabular}
\end{adjustbox}
\end{table}

\section{Supervised probing controls}
\label{app:supervised_controls}

To test whether the expert labels contain learnable, recording-dependent acoustic signal, we train lightweight heads on frozen audio representations; these controls are separate from the zero-shot evaluation in the main results.
Each recording is encoded with the frozen Qwen2.5-Omni-3B audio encoder and projection~\citep{qwen25omni} in 30-second windows with a 15-second hop, and projected audio-token states are mean-pooled into 2,048-dimensional vectors.
For Top-3 prediction, a linear head and an MLP ($2048\rightarrow256\rightarrow7$) applied to each window, with label-wise averaging of the top 20\% window logits, are trained on soft expert-selection targets with output biases initialized to the training-fold label prior.
We use five performer-grouped folds over 321 performer groups, with one fold for testing, the next for early stopping on development loss, and the remaining three for training.
For segment classification, heads are trained on expert-marked segments and evaluated on the 262 consensus events, holding out each of the nine source datasets in turn.
Metadata-only controls use the same objectives and splits with source dataset, subset, and duration features, adding relative position and segment length for segments.
Results average seeds 13, 37, and 73; differences use paired bootstrap intervals over performer groups (5,000 resamples).

On Top-3 prediction, the audio linear head exceeds the metadata-only control in macro-F1 by 6.79 points (95\% CI [4.99, 8.52]) and the audio MLP by 2.55 points (95\% CI [1.08, 4.06]), whereas neither differs significantly from the metadata-only control in F1@3 (Table~\ref{tab:supervised_controls}).
With source datasets held out, the segment MLP exceeds the metadata-only control in macro-F1 by 6.30 points (95\% CI [2.62, 10.03]) but remains below the majority baseline in any-match accuracy (30.92\% vs.\ 40.46\%).
These results indicate learnable, recording-dependent signal under class-balanced evaluation.

\begin{table}[H]
\centering
\caption{Supervised probing controls on frozen Qwen2.5-Omni-3B audio representations (\%), averaged over three seeds. The first metric column is F1@3 for Top-3 and any-match accuracy for segments.}
\label{tab:supervised_controls}
\small
\setlength{\tabcolsep}{6pt}
\begin{tabular}{@{}llcc@{}}
\toprule
\textbf{Task} & \textbf{Model} & \textbf{F1@3 / Match} & \textbf{Macro-F1} \\
\midrule
\multirow{4}{*}{\shortstack[l]{Top-3\\(5 performer-grouped folds)}} & Training-fold majority & 62.49 & 32.72 \\
 & Metadata-only linear & 63.19 & 41.74 \\
 & Audio linear & 62.70 & 48.52 \\
 & Audio MLP & 63.13 & 44.28 \\
\midrule
\multirow{4}{*}{\shortstack[l]{Segment\\(source dataset held out)}} & Other-source majority & 40.46 & 8.23 \\
 & Metadata-only linear & 30.66 & 14.13 \\
 & Audio linear & 25.95 & 20.08 \\
 & Audio MLP & 30.92 & 20.43 \\
\bottomrule
\end{tabular}
\end{table}

\section{Issue taxonomy guidance}
\label{app:taxonomy_guidance}

Table~\ref{tab:taxonomy_guidance} provides the category guidance shown to annotators for the seven fine-grained vocal-issue labels.
The English text below is translated from the Korean annotation guide.

\begin{table}[H]
\centering
\caption{Fine-grained vocal-issue taxonomy guidance.}
\label{tab:taxonomy_guidance}
\footnotesize
\setlength{\tabcolsep}{4pt}
\renewcommand{\arraystretch}{1.12}
\begin{tabularx}{\textwidth}{@{}llX@{}}
\toprule
\textbf{Parent} & \textbf{Label} & \textbf{Guidance} \\
\midrule
\multirow{3}{*}{Technical Production} & \textsc{Breath} & The overall use and control of breath, including phrase sustain, breath support, release of support during descending passages, and breath length or distribution. \\
 & \textsc{Vocalization} & Overall vocal quality and production method, including phonation, placement, blocked or constricted sound, muffled or pulled-back placement, high-note stability, and resonance. \\
 & \textsc{Technique} & The use and proficiency of special vocal skills, such as vibrato, vocal fry, bending, and ornamentation. \\
\midrule
\multirow{2}{*}{Musical Accuracy} & \textsc{Pitch} & All cases related to pitch accuracy and stability, including leaps, final-note stability, pitch sagging in descending passages, and the accuracy of ornaments or modulations. \\
 & \textsc{Rhythm} & Beat and timing, including rushing or dragging, pushing or pulling against the beat, and rhythmic expression through accent or dynamics. \\
\midrule
\multirow{2}{*}{Delivery} & \textsc{Diction} & Lyric intelligibility and pronunciation habits, including vowel maintenance, nasal tone, hard or blocked articulation, and genre-stylistic pronunciation. \\
 & \textsc{Expression} & Expressive capacity, including musical interpretation and emotional delivery. \\
\bottomrule
\end{tabularx}
\end{table}

\section{Audio source licenses}
\label{app:audio_source_licenses}

Table~\ref{tab:audio_source_licenses} summarizes the source datasets used in VocalCoachBench and their redistribution terms.
Track counts refer to the number of recordings included in VocalCoachBench, not the full size of the source corpus.

\begin{table}[H]
\centering
\caption{Audio source datasets and license terms.}
\label{tab:audio_source_licenses}
\footnotesize
\setlength{\tabcolsep}{4pt}
\renewcommand{\arraystretch}{1.08}
\begin{tabularx}{\textwidth}{@{}llrX@{}}
\toprule
\textbf{Subset} & \textbf{Source dataset} & \textbf{Tracks} & \textbf{License / terms} \\
\midrule
Same-song & DAMP-S-AG~\citep{damp_sag} & 207 & Smule Research Data License; research/education use only, no redistribution or commercial use \\
\midrule
Diverse-song & PopBuTFy~\citep{popbutfy} & 85 & CC BY-NC-SA 4.0 with dataset terms of access \\
Diverse-song & N20EMv2~\citep{n20emv2} & 65 & CC BY-SA 4.0 \\
Diverse-song & LM-SSD~\citep{LM-SSD} & 7 & CC BY 4.0 \\
Diverse-song & NUS-48E~\citep{nus48e} & 41 & Research-use-only corpus terms \\
Diverse-song & MedleyDB~\citep{medleydb} & 32 & CC BY-NC-SA 4.0 \\
Diverse-song & MedleyVox~\citep{medleyvox} & 12 & CC BY-NC-SA 4.0 \\
Diverse-song & MRSSing~\citep{mrssing} & 43 & CC BY 4.0 \\
Diverse-song & URSing~\citep{ursing} & 14 & CC BY 4.0 \\
Diverse-song & GTSinger~\citep{gtsinger} & 9 & CC BY-NC-SA 4.0 \\
\midrule
Total & & 515 & \\
\bottomrule
\end{tabularx}
\end{table}
Some source datasets do not permit redistribution of raw audio.
For these sources, we provide audio-access manifests rather than audio files. Benchmark annotations, metadata, prompts, and evaluation code are publicly released.
Redistributable audio is included where permitted by the original licenses.

\section{Public-data exposure sensitivity}
\label{app:exposure}

The diverse-song recordings come from public datasets, and most model training corpora are not fully documented, so prior exposure to these recordings cannot be audited directly.
As a sensitivity analysis, we remove the 32 recordings sourced from MedleyDB, whose multitracks are also redistributed in other public music corpora; 23 of them come from songs included in MUSDB18~\citep{musdb18}. The 12 MedleyVox recordings, which are also derived from MedleyDB tracks, are retained.
Removing these recordings changes each model's Top-3 macro-F1 by at most 1.34 points and segment macro-F1 by at most 0.54 points (Table~\ref{tab:exposure}).

\begin{table}[H]
\centering
\caption{Structured results (\%) on all recordings and after removing the 32 recordings sourced from MedleyDB (483 recordings and 241 segment events remain).}
\label{tab:exposure}
\scriptsize
\setlength{\tabcolsep}{3pt}
\renewcommand{\arraystretch}{1.08}
\begin{adjustbox}{max width=\textwidth}
\begin{tabular}{@{}llcccccc@{}}
\toprule
\textbf{Protocol} & \textbf{Model} & \textbf{\shortstack{F1@3\\All}} & \textbf{\shortstack{F1@3\\Excl.}} & \textbf{\shortstack{Macro-F1\\All}} & \textbf{\shortstack{Macro-F1\\Excl.}} & \textbf{\shortstack{Seg. macro-F1\\All}} & \textbf{\shortstack{Seg. macro-F1\\Excl.}} \\
\midrule
\multicolumn{2}{@{}l}{Majority (no audio)} & 62.49 & 63.11 & 32.72 & 32.92 & 8.23 & 8.02 \\
\midrule
\multirow{6}{*}{\textsc{Main}} & R1-AQA & 34.50 & 34.89 & 21.43 & 21.58 & 7.00 & 6.92 \\
 & Qwen2.5-Omni & 38.33 & 38.83 & 31.07 & 31.42 & 7.67 & 7.60 \\
 & Kimi-Audio & 33.82 & 33.85 & 28.25 & 28.16 & 13.63 & 13.52 \\
 & MiMo-Audio & 42.55 & 42.50 & 30.15 & 30.03 & 8.23 & 8.02 \\
 & Qwen3-Omni & 40.88 & 40.87 & 32.32 & 31.97 & 14.20 & 13.87 \\
 & Fun-Audio-Chat & 41.38 & 41.60 & 29.71 & 29.98 & 9.11 & 8.74 \\
\midrule
\multirow{4}{*}{\textsc{Closed}} & GPT-Audio 1.5 & 39.12 & 38.88 & 33.46 & 33.30 & 8.09 & 7.97 \\
 & Gemini 3 Flash Preview & 49.84 & 49.72 & 33.40 & 33.15 & 12.78 & 12.29 \\
 & Qwen3.5-Omni Flash & 44.54 & 44.66 & 31.61 & 31.39 & 14.65 & 14.12 \\
 & Qwen3.5-Omni Plus & 31.89 & 31.72 & 21.53 & 21.26 & 8.75 & 8.70 \\
\midrule
\multirow{2}{*}{\textsc{Simple}} & Audio Flamingo 3 & 33.67 & 34.07 & 23.28 & 23.62 & 13.53 & 13.51 \\
 & Music Flamingo & 43.46 & 43.34 & 33.32 & 31.98 & 3.82 & 4.12 \\
\bottomrule
\end{tabular}
\end{adjustbox}
\end{table}

\section{Atomic claim extraction}
\label{app:claim_extraction}

We decompose expert-written feedback blocks into atomic claims using an LLM with a fixed, meaning-preserving prompt.
The LLM works from the expert-written block text and its tags alone and decomposes it into normalized atomic units.
The prompt prohibits adding unsupported information, treats category tags only as interpretive hints, preserves original-language domain-specific terms, and only splits feedback into verifiable \texttt{issue}, \texttt{correction}, and \texttt{strength} claims.

We use the \texttt{gpt-5.4} API with deterministic decoding (\texttt{temperature=0}) and enforce JSON output with \texttt{response\_format=\{``type'': ``json\_object''\}}.
The system message is: \texttt{You output only JSON. No prose.} The main extraction is run with the OpenAI Batch API over 4,506 feedback blocks.
Each correction claim includes a \texttt{targets} field linking it to the issue claim(s) it addresses, or an empty list when no issue is explicitly or implicitly targeted.

Because claim decomposition may involve minimal normalization, such as restoring omitted subjects or linking corrections to issue claims, the resulting claims should be understood as a structured representation of human expert feedback rather than independent expert annotations.
We mitigate this limitation by using deterministic decoding, a fixed source-faithful prompt, explicit grounding constraints, and by reporting the prompt and model settings below and releasing the final extracted claims.

\paragraph{Claim extraction prompt.}
The following is an English-rendered version of the claim extraction prompt.
The prompt used for extraction includes Korean lexical examples and domain terms. The placeholders \texttt{\{\{CATEGORIES\}\}} and \texttt{\{\{BLOCK\_TEXT\}\}} are filled with the rubric tags and Korean feedback text for each block.

\noindent\textbf{Task and input.}
\begin{promptbox}
You are an annotator who decomposes Korean vocal coaching feedback blocks into atomic coaching claims.\\[2pt]
Input: categories are rubric tags assigned to the block and should be used only as hints; block\_text is Korean feedback written by a vocal coaching annotator.\\[2pt]
Important constraint: categories are only hints for interpretation. Do not create any claim that is not directly supported by block\_text.
\end{promptbox}

\noindent\textbf{Claim schema.}
\begin{promptbox}
Use only one of the following three claim types.\\[2pt]
issue: a negative observation or problem in the singer's performance.\\[2pt]
correction: a concrete corrective action or practice method. A correction must specify how to improve; do not extract vague advice such as ``practice more'', ``pay attention'', or ``improve this''.\\[2pt]
strength: a positive observation or well-performed aspect of the singer's performance.\\[4pt]
Each claim must include id, type, and text. The id is unique within the block and follows the order of appearance: ``c1'', ``c2'', ... . The text is one self-contained Korean sentence.
Correction claims additionally include targets, a list of issue IDs addressed by the correction.
Do not include targets for issue or strength claims.
If a correction does not address any issue explicitly or implicitly present in the block, use ``targets'': [].
\end{promptbox}

\noindent\textbf{Normalization rules.}
\begin{promptbox}
1. One claim must contain one verifiable statement.\\[2pt]
2. Remove unnecessary discourse connectives and deictic expressions.\\[2pt]
3. Restore omitted subjects or pronouns only when they are clear from block\_text itself, not from categories.\\[2pt]
4. Do not add information, interpretation, severity, or explanation that is not supported by the original text.\\[2pt]
5. Do not summarize, expand, soften, or intensify the meaning.\\[2pt]
6. Remove final periods.\\[2pt]
7. Keep Korean. Do not translate claim text into English.\\[2pt]
8. Preserve domain-specific terms when they appear in the original text.
\end{promptbox}

\noindent\textbf{Splitting and correction rules.}
\begin{promptbox}
1. If cause and effect appear in one statement, keep them as one issue.\\[2pt]
2. If different issues appear in one sentence, split them into separate issue claims.\\[2pt]
3. If multiple practice methods target the same issue, combine them into one correction claim.\\[2pt]
4. If multiple practice methods target different issues, split them into separate correction claims and connect each one to its target issue.\\[2pt]
5. Do not extract metaphors or examples as independent claims. If a metaphor supports a nearby correction, absorb only the practical instruction into that correction; otherwise discard it.\\[2pt]
6. If the block contains a correction but no explicit issue, create an implied issue only when the problem state is directly inferable from the correction sentence itself. Do not create an issue based only on categories.\\[2pt]
7. If there are no extractable issue, correction, or strength claims, return an empty claims list.\\[2pt]
8. Treat restatement and elaboration as part of the original claim, not as a new claim. Merge additional specificity into the existing claim text instead of producing duplicates.\\[4pt]
Extract a correction only when it includes a concrete behavior, practice method, vocal adjustment, pronunciation method, body/breath control method, or tool. Do not extract vague or how-less advice as correction.
\end{promptbox}

\noindent\textbf{Output format.}
\begin{promptbox}
Return JSON only. Do not include explanations, markdown, or any text outside JSON.\\[4pt]
\{\\
\hspace*{1em}``claims'': [\\
\hspace*{2em}\{``id'': ``c1'', ``type'': ``issue'', ``text'': ``...''\},\\
\hspace*{2em}\{``id'': ``c2'', ``type'': ``correction'', ``text'': ``...'', ``targets'': [``c1'']\}\\
\hspace*{1em}]\\
\}\\[4pt]
If there are no extractable claims, return: \{ ``claims'': [] \}\\[4pt]
Block to process:\\
categories: \{\{CATEGORIES\}\}\\
block\_text: \{\{BLOCK\_TEXT\}\}
\end{promptbox}

\subsection{Single-author audit of claim extraction}
\label{app:claim_audit}

To validate the atomic-claim extraction stage, one author manually audited 100 source feedback blocks sampled from the full set of 4,506 blocks. 
We used stratified sampling over extracted claim-type composition, preserving the relative mixture of blocks containing \emph{diagnosis}, \emph{correction}, and \emph{strength} claims. 
For each sampled block, the author reviewed the original Korean feedback text alongside the extracted atomic claims.

\begin{table}[h]
\centering
\small
\begin{tabular}{lc}
\toprule
Error type & Rate \\
\midrule
Atomicity violation & 0.8\% \\
Fragmentation error & 0.0\% \\
Omission & 0.0\% \\
Hallucinated addition & 0.0\% \\
Contextual role-assignment error & 0.4\% \\
Correction target-link error & 0.0\% \\
\bottomrule
\end{tabular}
\caption{Single-author audit results for claim extraction. Correction target-link errors are computed over audited correction claims; other error rates are computed over audited extracted claims.}
\label{tab:claim_audit}
\end{table}

The observed role-assignment errors were context-sensitive boundary cases rather than unsupported additions: the extracted text was grounded in the original Korean block, but its role as a diagnosis, correction, or strength was occasionally misread.

\section{Judge validation details}
\label{app:judge_validation}

We validate the open-ended evaluation judges on label-balanced samples from model-evaluation outputs.
These validation samples are designed to stress rubric boundaries rather than estimate natural label prevalence.

\subsection{Diagnosis judge validation}
\label{app:diagnosis_judge_validation}

We validate diagnosis-side judge labels using two author relabelings of the same 100 cases, sampled as 25 cases from each original LLM judge label: \emph{strict hit}, \emph{coarse hit}, \emph{miss}, and \emph{contradict}.
Cases marked \texttt{BAD\_CASE} or left blank are excluded from the main agreement calculations.

\begin{table}[H]
\centering
\caption{Diagnosis judge validation summary. Exact agreement uses the four-way diagnosis labels. Binary agreement collapses \textsc{Strict Hit} and \textsc{Coarse Hit} into \textsc{Hit}, and \textsc{Miss} and \textsc{Contradict} into \textsc{Non-Hit}.}
\label{tab:diagnosis_judge_validation_summary}
\footnotesize
\setlength{\tabcolsep}{5pt}
\renewcommand{\arraystretch}{1.08}
\begin{tabular}{@{}lccc@{}}
\toprule
\textbf{Comparison} & \textbf{Valid cases} & \textbf{Agreement} & \textbf{Cohen's $\kappa$} \\
\midrule
LLM judge vs.\ Annotator 1 & 99 & 84.8 & .798 \\
LLM judge vs.\ Annotator 2 & 94 & 81.9 & .759 \\
LLM judge vs.\ authors, pooled & 193 & 83.4 & .779 \\
Annotator 1 vs.\ Annotator 2, exact & 93 & 80.6 & .737 \\
Annotator 1 vs.\ Annotator 2, binary & 93 & 93.5 & .870 \\
\bottomrule
\end{tabular}
\end{table}

\begin{table}[H]
\centering
\caption{Pooled four-way confusion matrix for diagnosis judge validation. Rows are LLM judge labels and columns are author labels, after excluding \texttt{BAD\_CASE} and blank labels.}
\label{tab:diagnosis_judge_confusion_pooled}
\footnotesize
\setlength{\tabcolsep}{5pt}
\renewcommand{\arraystretch}{1.12}
\begin{tabular}{@{}lcccc@{}}
\toprule
\textbf{LLM judge $\backslash$ author} & \textbf{Strict} & \textbf{Coarse} & \textbf{Miss} & \textbf{Contra.} \\
\midrule
Strict & \cellcolor{green!35}48 & \cellcolor{orange!10}2 & 0 & 0 \\
Coarse & \cellcolor{orange!25}13 & \cellcolor{green!28}33 & \cellcolor{orange!10}2 & 0 \\
Miss & 0 & 0 & \cellcolor{green!35}47 & \cellcolor{orange!8}1 \\
Contra. & \cellcolor{orange!8}1 & \cellcolor{orange!14}5 & \cellcolor{orange!20}8 & \cellcolor{green!28}33 \\
\bottomrule
\end{tabular}
\end{table}

\begin{table}[H]
\centering
\caption{Author--author four-way confusion matrix for diagnosis validation}
\label{tab:diagnosis_author_confusion}
\footnotesize
\setlength{\tabcolsep}{5pt}
\renewcommand{\arraystretch}{1.12}
\begin{tabular}{@{}lcccc@{}}
\toprule
\textbf{Annotator 2 $\backslash$ Annotator 1} & \textbf{Strict} & \textbf{Coarse} & \textbf{Miss} & \textbf{Contra.} \\
\midrule
Strict & \cellcolor{green!28}26 & \cellcolor{orange!10}2 & 0 & 0 \\
Coarse & \cellcolor{orange!20}7 & \cellcolor{green!18}12 & \cellcolor{orange!10}2 & \cellcolor{orange!10}2 \\
Miss & 0 & \cellcolor{orange!10}2 & \cellcolor{green!28}24 & \cellcolor{orange!12}3 \\
Contra. & 0 & 0 & 0 & \cellcolor{green!18}13 \\
\bottomrule
\end{tabular}
\end{table}

\subsection{Correction judge validation}
\label{app:correction_judge_validation}

For correction appropriateness, a professional vocal coach reviewed a 100-case correction-level validation set. 
Each case included the model diagnosis, model prescription, the matched expert issue, and the linked expert correction.
Of the 100 cases, 97 received usable expert labels, one was marked as an unusable case, and two remained unlabeled.

The main correction metric uses three labels: \emph{valid}, \emph{weak}, and \emph{invalid}.
After excluding auxiliary safety-diagnostic cases and unusable/unlabeled cases, the three-label validation set contains 72 correction-level decisions.
Agreement between the LLM judge and the vocal coach is 79.2\% exact agreement with Cohen's $\kappa=0.687$.

\begin{table}[H]
\centering
\caption{
Correction judge validation confusion matrix for the three main correction labels. Rows are LLM judge labels and columns are expert labels.
}
\label{tab:correction_judge_validation_confusion}
\small
\setlength{\tabcolsep}{8pt}
\renewcommand{\arraystretch}{1.12}
\begin{tabular}{@{}lrrrr@{}}
\toprule
\textbf{LLM judge / Expert} & \textbf{Valid} & \textbf{Weak} & \textbf{Invalid} & \textbf{Total} \\
\midrule
Valid   & 21 & 3  & 0  & 24 \\
Weak    & 5  & 19 & 1  & 25 \\
Invalid & 4  & 2  & 17 & 23 \\
\midrule
Total   & 30 & 24 & 18 & 72 \\
\bottomrule
\end{tabular}
\end{table}

Most disagreements occur on the boundary between \emph{valid} and \emph{weak}, reflecting whether a correction is judged sufficiently specific to the target issue or merely relevant but underspecified.
The \emph{invalid} label is more stable, with 17 of 23 LLM-judged invalid cases also labeled invalid by the expert.

\paragraph{Auxiliary safety diagnostic.}
We initially included a separate safety-oriented label for corrections that could encourage harmful vocal practice.
Because such cases were rare and showed lower boundary agreement in the small validation audit, we exclude this label from the main correction metrics and retain it only as an auxiliary diagnostic flag.
As a binary diagnostic, this flag achieved 84.5\% agreement with Cohen's $\kappa=0.482$ on the 97 usable expert-labeled cases.

\subsection{Judge sensitivity}
\label{app:judge_sensitivity}

The human audits above do not test whether model-level results depend on the judge prompt or judge model.
We therefore re-judged a model-stratified sample of 1,200 diagnosis and 1,200 correction cases, 100 per model, drawn from outputs of an earlier evaluation run of the 12 models, so the analysis characterizes judge behavior rather than re-scoring Table~\ref{tab:open_ended_coaching_results}.
Each case was judged by GPT-OSS-120B and Claude Sonnet 5, each with the original rubric prompt and a meaning-preserving paraphrase (Table~\ref{tab:judge_sensitivity}).

A paraphrased prompt leaves most decisions unchanged, with binary agreement of 90.5\% for diagnosis and 92.4\% for correction under GPT-OSS-120B.
Across judge models, diagnosis decisions agree on 87.9\% of cases and the top Main model is unchanged.
Strict correction validity is judge-dependent: Claude Sonnet 5 labels 26.1\% of cases \emph{valid} compared with 72.6\% for GPT-OSS-120B, binary agreement is 52.0\%, the Main-model Spearman correlation is 0.706, and the top Main model changes.
Agreement on whether a prescription is \emph{valid} or \emph{weak} is 90.8\%.
We therefore interpret strict prescription validity as a judge-calibrated measure of specificity and read it together with the \emph{weak} rate.

\begin{table}[H]
\centering
\caption{Judge sensitivity on 1,200 diagnosis and 1,200 correction cases (100 per model) from outputs of an earlier evaluation run of the 12 models. Binary decisions are strict-or-coarse hit for diagnosis and \emph{valid} for correction. $\rho$ is the Spearman correlation of binary rates across the six Main models; Top asks whether the highest-rated Main model is the same.}
\label{tab:judge_sensitivity}
\small
\setlength{\tabcolsep}{5pt}
\begin{tabular}{@{}llccccc@{}}
\toprule
\textbf{Task} & \textbf{Comparison} & \textbf{Exact} & \textbf{$\kappa$} & \textbf{Binary} & \textbf{Binary $\kappa$} & \textbf{$\rho$ / Top} \\
\midrule
\multirow{3}{*}{Diagnosis} & Prompt paraphrase, GPT-OSS-120B & 84.2 & 0.741 & 90.5 & 0.810 & 0.943 / same \\
 & Prompt paraphrase, Claude Sonnet 5 & 92.2 & 0.858 & 93.6 & 0.871 & 0.943 / same \\
 & Judge model, original prompts & 75.8 & 0.588 & 87.9 & 0.758 & 0.943 / same \\
\midrule
\multirow{3}{*}{Correction} & Prompt paraphrase, GPT-OSS-120B & 88.8 & 0.743 & 92.4 & 0.809 & 0.986 / same \\
 & Prompt paraphrase, Claude Sonnet 5 & 88.8 & 0.774 & 94.0 & 0.842 & 0.868 / same \\
 & Judge model, original prompts & 43.5 & 0.199 & 52.0 & 0.211 & 0.706 / differs \\
\bottomrule
\end{tabular}
\end{table}

\section{Prompt templates}
\label{app:prompts}

This appendix lists the prompts used by the prompting protocols described in Section~\ref{sec:prompting_inference}.
All prompts are audio-only: they do not include sample identifiers, dataset labels, filenames, or song-title metadata in the model instruction.
The seven allowed category IDs (\texttt{PITCH, RHYTHM, DICTION, BREATH, PHONATION, TECHNIQUE, EXPRESSION}) follow the taxonomy in Section~\ref{sec:annotation_protocol}; the runtime token \texttt{PHONATION} is normalized to the human-facing vocalization category.

\subsection{Claim/objective protocol (main setting)}
\label{app:prompts_claimobj}

The main protocol decomposes evaluation into three task-specific prompts: a coaching-claim prompt, an objective Top-3 + score prompt, and a structured segment prompt.

\noindent\textbf{Coaching-claim prompt}
\begin{promptbox}
Listen to the singing audio as a vocal coach.\\[2pt]
Categories: PITCH, RHYTHM, DICTION, BREATH, PHONATION, TECHNIQUE, EXPRESSION.\\[2pt]
Return one valid JSON object only. Start with \{ and end with \}. Quote all keys and strings. Do not use markdown or YAML.\\[2pt]
Required keys: diagnosis\_claims (5 objects with id d1--d5, category, text, confidence), prescription\_claims (5 objects with id p1--p5, target\_diagnosis\_id, category, text, confidence), and strength\_claims (1 object with id s1, category, text, confidence).\\[2pt]
Rules: p1 targets d1, p2 targets d2, p3 targets d3, p4 targets d4, p5 targets d5. Each diagnosis names one specific vocal problem. Each prescription gives one concrete exercise or adjustment for its mapped diagnosis. Each strength names one audible vocal strength. Confidence is 0.0 to 1.0. Choose categories from the audio, not list order. Keep text concise.
\end{promptbox}

\noindent\textbf{Objective Top-3 + score prompt}
\begin{promptbox}
Evaluate the singing audio.\\[2pt]
Categories: PITCH, RHYTHM, DICTION, BREATH, PHONATION, TECHNIQUE, EXPRESSION.\\[2pt]
Return one valid JSON object only. Start with \{ and end with \}. Quote all keys and strings. Do not use markdown or YAML.\\[2pt]
Required keys: top3\_issues (exactly 3 category IDs, worst first), top3\_issue\_confidences (3 numbers from 0.0 to 1.0, aligned with top3\_issues), quality\_score\_0\_5 (one number from 0.0 to 5.0), score\_confidence (one number from 0.0 to 1.0), and score\_rationale (one short reason based on the singing).\\[2pt]
Score guide: 0 impossible, 1 very weak, 2 many major problems, 3 adequate amateur, 4 good with minor issues, 5 excellent.
Choose categories and score from the audio, not list order or defaults.
\end{promptbox}

\noindent\textbf{Structured segment prompt}
\begin{promptbox}
Evaluate this short problematic singing segment.\\[2pt]
Categories: PITCH, RHYTHM, DICTION, BREATH, PHONATION, TECHNIQUE, EXPRESSION.\\[2pt]
Return one valid JSON object only. Start with \{ and end with \}. Quote all keys and strings. Do not use markdown or YAML.\\[2pt]
Required keys: category (one category ID), confidence (one number from 0.0 to 1.0), and rationale (one short reason based on the segment).\\[2pt]
Choose the main issue from the segment, not list order.
\end{promptbox}

\subsection{Simple protocol}
\label{app:prompts_simple}

The simple protocol replaces each structured prompt with a single-sentence natural-language instruction without any JSON constraint.
This protocol is used for models with limited support for long, multi-section schema-constrained instructions (Section~\ref{sec:models}).

\noindent\textbf{Simple overall coaching prompt}
\begin{promptbox}
You are receiving a singing audio file with this instruction, so listen to the audio and give short vocal coaching feedback in English that covers five main singing problems, five matching ways the singer can improve those problems in the same order, and one thing the singer does well.
\end{promptbox}

\noindent\textbf{Simple Top-3 prompt}
\begin{promptbox}
You are receiving a singing audio file with this instruction, so listen to the audio and choose exactly three biggest singing problem categories from PITCH, RHYTHM, DICTION, BREATH, PHONATION, TECHNIQUE, and EXPRESSION; answer in order from the most serious problem to the third most serious problem, choose from the audio rather than the category list order, and answer with only the three category names.
\end{promptbox}

\noindent\textbf{Simple score prompt}
\begin{promptbox}
You are receiving a singing audio file with this instruction, so listen to the audio and answer with only one number from 0 to 5 for overall singing quality, where 0 is impossible to evaluate, 1 is very weak, 2 has many problems, 3 is average amateur, 4 is good, and 5 is excellent.
\end{promptbox}

\medskip
\noindent\textbf{Simple segment prompt}
\begin{promptbox}
You are receiving a short problematic singing segment with this instruction, so listen to the segment and answer with exactly one main problem category from PITCH, RHYTHM, DICTION, BREATH, PHONATION, TECHNIQUE, and EXPRESSION; do not choose by list order.
\end{promptbox}

\subsection{Direct pairwise comparison}
\label{app:direct_prompt}
The canonical Direct task supplies labeled Audio A and Audio B:
\begin{promptbox}
Compare two singing audios as a vocal coach.\\[2pt]
You will receive two labeled audio inputs: Audio A and Audio B.
Audio A and Audio B are different performances of the same singing task. Choose which performance is better overall for vocal quality and musical execution.\\[2pt]
Return one valid JSON object only. Start with \{ and end with \}. Quote all keys and strings. Do not use markdown or YAML.\\[2pt]
Required keys: winner (``A'' or ``B''), confidence (one number from 0.0 to 1.0), and rationale (one short audio-grounded reason for the choice).\\[2pt]
Judge the relative performance quality. If the performances are close, still choose the slightly better one.
\end{promptbox}
Audio Flamingo 3 and Music Flamingo instead receive a single waveform containing A, one second of silence, and B, with interval instructions identifying each performance.

\subsection{Inference settings}
\label{app:inference_settings}

All models are queried with deterministic decoding when supported, using temperature 0 and no sampling. 
For Kimi-Audio, text decoding uses temperature 0, and the remaining settings follow the model authors' recommended configuration: audio temperature 0.8, audio top-$k$ 10, and text top-$k$ 5. 
All inference runs are executed with resumption-enabled batching to tolerate interruptions during large multi-model evaluation.
For API models, results come from the stored responses of past API runs; these runs do not establish equivalence of provider-internal model snapshots or of retries before the stored response.

Table~\ref{tab:inference_budgets} lists the generation budgets, which are larger for open-ended coaching than for label prompts.

\begin{table}[h]
\centering
\caption{Generation budgets used for model inference.}
\label{tab:inference_budgets}
\small
\setlength{\tabcolsep}{5pt}
\renewcommand{\arraystretch}{1.08}
\begin{tabular}{llr}
\toprule
\textbf{Protocol} & \textbf{Prompt / task} & \textbf{Max tokens} \\
\midrule
Main & Open-ended coaching feedback & 1,200 \\
Main & Objective Top-3 issue prediction & 420 \\
Main & Expert-localized segment issue classification & 256 \\
Simple & Overall coaching review & 650 \\
Simple & Short-form label prompts & 40--80 \\
Direct & Pairwise comparison & 480 \\
\bottomrule
\end{tabular}
\end{table}

\subsection{Open-ended response diversity diagnostics}
\label{app:open_ended_diversity}

Open-ended coaching metrics can be inflated when a model repeatedly emits broad, reusable feedback templates that partially overlap with common expert claims.
We therefore report response-diversity diagnostics as a companion analysis to Table~\ref{tab:open_ended_coaching_results}.
These diagnostics indicate whether a model's diagnosis and correction claims vary across recordings or collapse to repeated category and claim patterns.

For each model, we normalize extracted diagnosis and correction claim texts by lowercasing, removing punctuation, stripping extra whitespace, and collapsing exact duplicate normalized strings.
Diag.\ unique and Corr.\ unique report the percentage of unique normalized claim texts among all extracted diagnosis and correction claims.
Cat.\ H is normalized entropy over the seven diagnosis categories.
Canon-5 measures the fraction of outputs whose diagnosis category set exactly matches \{PITCH, RHYTHM, DICTION, BREATH, VOCALIZATION\}, a common list-order template observed in several models.
Top diag.\ rep.\ and Top corr.\ rep.\ report how often the single most repeated diagnosis or correction claim appears among non-empty outputs.
Higher unique rates and category entropy suggest more varied responses, while lower Canon-5 and repeat rates suggest less template collapse.

\begin{table}[H]
\centering
\caption{
Response-diversity diagnostics for open-ended coaching outputs. Unique rates count distinct normalized claim texts among diagnosis and correction claims. Cat.\ H is normalized diagnosis-category entropy over the seven issue labels. Canon-5 is the fraction of responses whose diagnosis categories exactly match \{PITCH, RHYTHM, DICTION, BREATH, VOCALIZATION\}; Top diag.\ and Top corr.\ repeat rates count how often the most repeated claim text appears among non-empty outputs.
}
\label{tab:open_ended_diversity}
\scriptsize
\setlength{\tabcolsep}{3.0pt}
\renewcommand{\arraystretch}{1.08}
\begin{adjustbox}{max width=\textwidth}
\begin{tabular}{@{}lcccccc@{}}
\toprule
\textbf{Model} &
\textbf{Diag. unique $\uparrow$} &
\textbf{Corr. unique $\uparrow$} &
\textbf{Cat. H $\uparrow$} &
\textbf{Canon-5 $\downarrow$} &
\textbf{Top diag. rep. $\downarrow$} &
\textbf{Top corr. rep. $\downarrow$} \\
\midrule
\rowcolor{gray!8}
\multicolumn{7}{l}{\textit{Main (claim/objective) protocol -- open-weight models}} \\
R1-AQA                 & 0.2 & 0.2 & 0.83 & 100.0 & 100.0 & 100.0 \\
Qwen2.5-Omni      & 7.5 & 18.8 & 0.87 & 93.8 & 46.6 & 37.9 \\
Kimi-Audio        & 43.0 & 53.2 & 0.87 & 30.3 & 27.0 & 14.6 \\
MiMo-Audio        & 33.8 & 45.8 & 0.88 & 0.0 & 13.5 & 12.7 \\
Qwen3-Omni         & 62.8 & 59.0 & 0.91 & 72.8 & 11.7 & 16.5 \\
Fun-Audio-Chat & 17.8 & 35.1 & 0.93 & 62.7 & 28.8 & 17.5 \\
\rowcolor{gray!8}
\multicolumn{7}{l}{\textit{Main (claim/objective) protocol -- closed-source API models}} \\
GPT-Audio 1.5      & 20.4 & 24.8 & 0.85 & 0.0 & 22.5 & 20.2 \\
Gemini 3 Flash Preview & 34.3 & 57.0 & 0.91 & 34.0 & 37.4 & 12.1 \\
Qwen3.5-Omni Flash & 86.2 & 98.3 & 0.89 & 71.8 & 14.3 & 1.2 \\
Qwen3.5-Omni Plus  & 75.1 & 89.6 & 0.87 & 8.7 & 6.8 & 3.7 \\
\rowcolor{gray!8}
\multicolumn{7}{l}{\textit{Simple protocol -- open-weight models with limited support for long structured prompts}} \\
Audio Flamingo 3 & 29.8 & 51.3 & 0.93 & 0.2 & 29.5 & 20.1 \\
Music Flamingo    & 58.9 & 70.8 & 0.85 & 0.0 & 28.3 & 18.8 \\
\bottomrule
\end{tabular}
\end{adjustbox}
\end{table}

R1-AQA almost completely collapses to a fixed template (Table~\ref{tab:open_ended_diversity}): both unique rates are 0.2\%, Canon-5 is 100.0\%, and the top repeated diagnosis and correction claims appear in every non-empty output.
Qwen2.5-Omni and Fun-Audio-Chat also show strong template effects, with high Canon-5 rates and low unique rates.
This suggests that part of their open-ended coverage may come from broad, reusable coaching statements rather than fully audio-specific feedback.
In contrast, Kimi-Audio, MiMo-Audio, Qwen3-Omni, Gemini 3 Flash Preview, and the Qwen3.5-Omni models generally produce more varied claim text, whereas GPT-Audio 1.5 repeats claim text more often (unique rates of 20.4\% and 24.8\%) without following the Canon-5 template.
However, high surface diversity alone is not sufficient: for example, Qwen3.5-Omni Flash has very high unique rates but still frequently follows the Canon-5 category template.
We therefore interpret diversity diagnostics jointly with the LLM-judge metrics: high diagnosis hit or correction validity is more convincing when accompanied by low repetition and low template-match rates.

\subsection{Post-processing pipeline}
\label{app:postprocessing}

Raw model outputs pass through protocol-specific validation and normalization before metric computation.
\emph{(i) Direct structured parsing}
For the claim, objective, and structured segment prompts, models are instructed to emit JSON directly; outputs are parsed against the required fields, and generations from which no required field or label can be recovered are counted as format failures.
\emph{(ii) Simple-prompt normalization}
For simple-prompt outputs (Top-3 category list, quality score, and segment label), deterministic text parsers map surface text to canonical taxonomy labels.
\emph{(iii) Claim extraction and judging}
Free-form simple-protocol overall reviews are first decomposed by GPT-OSS-120B~\citep{gptoss} into structured diagnosis and correction claims, after which the diagnosis and correction judges compute the open-ended metrics for all samples.

\section{Diagnosis-conditioned correction}
\label{app:conditioned_correction}

Table~\ref{tab:conditioned_correction} relates prescription validity to whether the model's own target diagnosis is supported by the expert references.
A prescription is supported when its linked model diagnosis matches at least one expert diagnosis claim with a strict or coarse hit; strict support requires a strict hit.
Links follow the end-to-end metric, including the positional fallback for Simple outputs.

For the closed-source models, validity differs by at most 2.05 points between supported and unsupported targets, so their high validity does not depend on diagnosis alignment.
Requiring strict support lowers end-to-end rates to at most 14.10\% (Qwen3.5-Omni Plus).

\begin{table}[H]
\centering
\caption{Diagnosis-conditioned correction (\%). Valid and E2E divide by the 2,575 fixed prescription slots; E2E (strict) additionally requires a strict hit. The last two columns give the \emph{valid} rate among emitted prescriptions whose linked model diagnosis is, or is not, supported by an expert diagnosis claim through a strict or coarse hit.}
\label{tab:conditioned_correction}
\scriptsize
\setlength{\tabcolsep}{3pt}
\renewcommand{\arraystretch}{1.08}
\begin{adjustbox}{max width=\textwidth}
\begin{tabular}{@{}llcccccc@{}}
\toprule
\textbf{Protocol} & \textbf{Model} & \textbf{Emitted} & \textbf{Valid} & \textbf{E2E} & \textbf{\shortstack{E2E\\(strict)}} & \textbf{\shortstack{Valid $|$\\supported}} & \textbf{\shortstack{Valid $|$\\unsupported}} \\
\midrule
\multirow{6}{*}{\textsc{Main}} & R1-AQA & 2,575 & 24.50 & 8.62 & 1.32 & 49.33 & 19.25 \\
 & Qwen2.5-Omni & 2,575 & 29.48 & 19.42 & 2.10 & 28.77 & 30.94 \\
 & Kimi-Audio & 925 & 19.61 & 12.39 & 1.44 & 53.43 & 56.71 \\
 & MiMo-Audio & 2,565 & 84.70 & 42.68 & 5.20 & 86.60 & 83.49 \\
 & Qwen3-Omni & 2,575 & 88.39 & 53.67 & 10.87 & 93.06 & 82.02 \\
 & Fun-Audio-Chat & 2,485 & 66.68 & 47.07 & 7.30 & 73.14 & 60.99 \\
\midrule
\multirow{4}{*}{\textsc{Closed}} & GPT-Audio 1.5 & 2,575 & 92.58 & 57.55 & 8.23 & 92.80 & 92.23 \\
 & Gemini 3 Flash Preview & 2,565 & 78.29 & 33.13 & 2.99 & 79.79 & 77.74 \\
 & Qwen3.5-Omni Flash & 2,560 & 94.72 & 58.02 & 12.85 & 95.59 & 94.78 \\
 & Qwen3.5-Omni Plus & 2,555 & 96.85 & 63.96 & 14.10 & 97.86 & 97.13 \\
\midrule
\multirow{2}{*}{\textsc{Simple}} & Audio Flamingo 3 & 1,276 & 5.44 & 2.80 & 0.43 & 20.99 & 7.29 \\
 & Music Flamingo & 514 & 4.58 & 0.89 & 0.12 & 58.97 & 20.00 \\
\bottomrule
\end{tabular}
\end{adjustbox}
\end{table}

\clearpage
\section{Annotation process}
\label{app:annotation_interface}

\subsection{Annotation interface}
Figures show the full annotation interfaces used by expert annotators. The screenshots preserve the original Korean interface text used during annotation.
\begin{figure}[H]
\centering
\includegraphics[width=0.88\textwidth,height=0.36\textheight,keepaspectratio]{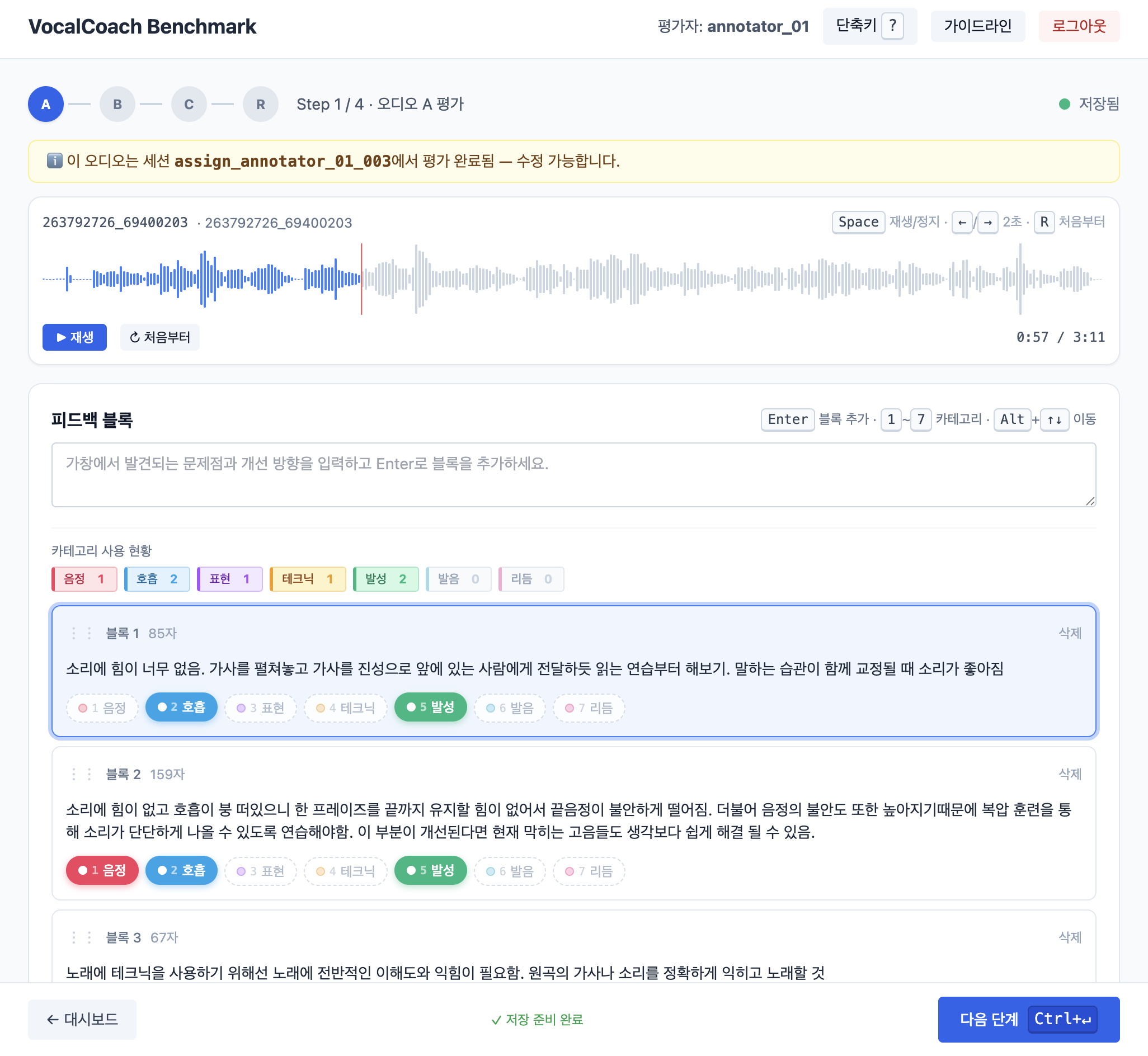}

\vspace{0.5em}

\includegraphics[width=0.88\textwidth,height=0.36\textheight,keepaspectratio]{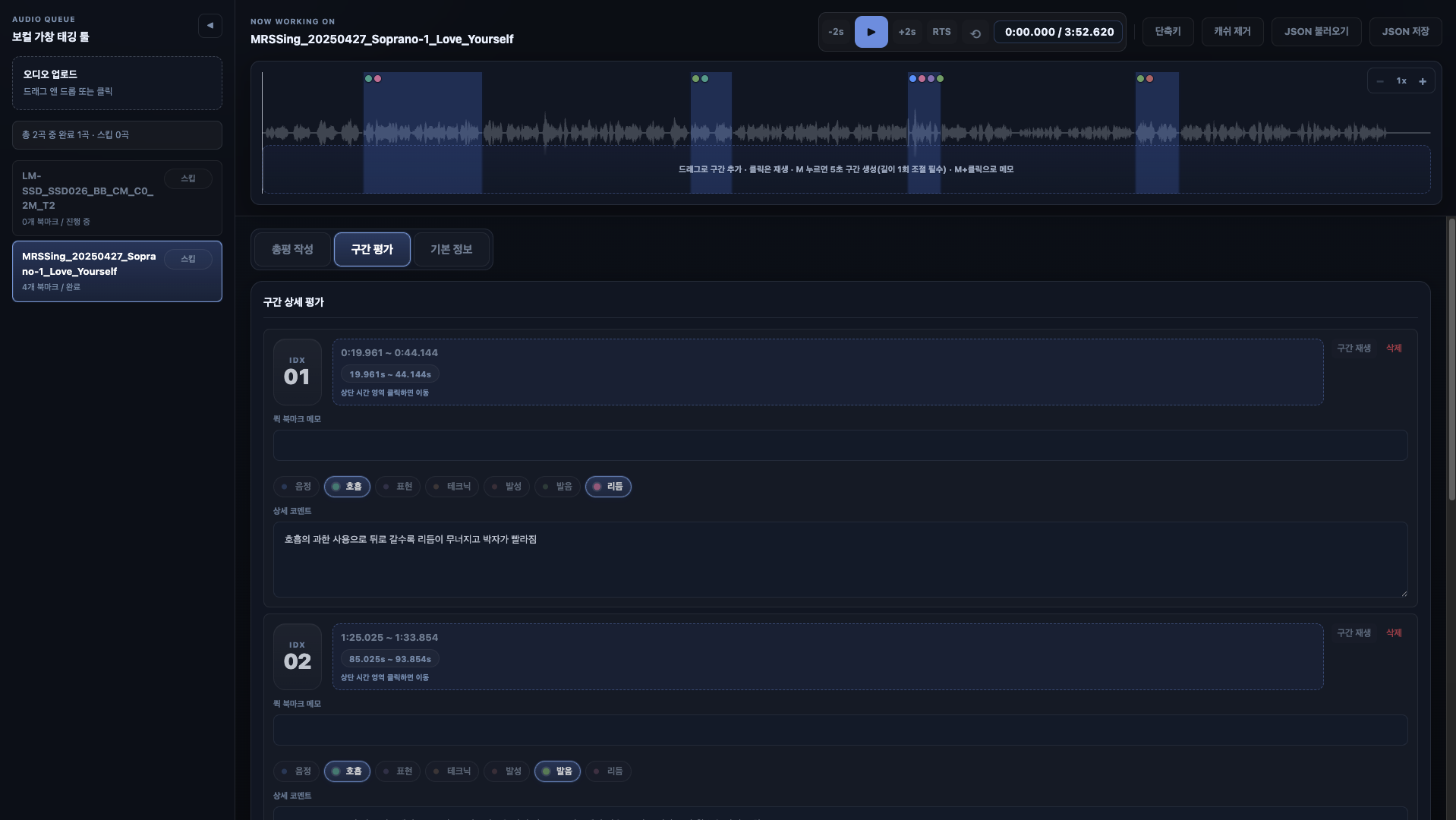}
\caption{
Annotation interfaces. Top: same-song interface for holistic coaching feedback. Bottom: diverse-song interface for segment-level issue annotation.
}
\label{fig:annotation_ui}
\end{figure}

\subsection{Annotator training, compensation, and quality control}

All annotators were compensated at an hourly rate equivalent to approximately USD 20, and the overall annotation process lasted approximately seven days. Prior to annotation, annotators attended a training session covering the task objectives and annotation guidelines.

For calibration and quality control, the workload was gradually increased: annotators completed two examples on the first day and five examples on the second day. After these initial annotations were reviewed for consistency with the guidelines, the main annotation phase began on the third day and continued for four days. To prevent fatigue and maintain annotation quality, annotators were allowed to submit only a fixed amount of data per day, corresponding to approximately 1.5 hours of work.

\section{Compute resources}
\label{app:compute_resources}

We do not train or fine-tune the evaluated audio-language models; the supervised controls in Appendix~\ref{app:supervised_controls} train only small heads on frozen audio representations. Open-weight model inference was run on a single NVIDIA H100 80GB HBM3 GPU (CUDA 12.8), with one model evaluated at a time. Closed-source models were evaluated through public APIs, and LLM-based parsing, normalization, and judging were run as resumable batch jobs. Wall-clock time varied substantially by model and API latency, so the released scripts support checkpointing and resumption.

\clearpage

\clearpage
\section*{NeurIPS Paper Checklist}

\begin{enumerate}

\item {\bf Claims}
    \item[] Question: Do the main claims made in the abstract and introduction accurately reflect the paper's contributions and scope?
    \item[] Answer: \answerYes{}.
    \item[] Justification: The abstract and introduction state the benchmark, annotation, and evaluation contributions, and the Results and Limitations sections qualify the scope of the empirical claims.

\item {\bf Limitations}
    \item[] Question: Does the paper discuss the limitations of the work performed by the authors?
    \item[] Answer: \answerYes{}.
    \item[] Justification: The paper includes a Limitations and future work paragraph discussing the English solo-singing scope, single-song same-song subset, audio-only setting, expert-provided segments, and limitations of reference-supported open-ended evaluation.

\item {\bf Theory assumptions and proofs}
    \item[] Question: For each theoretical result, does the paper provide the full set of assumptions and a complete (and correct) proof?
    \item[] Answer: \answerNA{}.
    \item[] Justification: The paper introduces a benchmark and empirical evaluation suite and does not include theoretical results or proofs.

    \item {\bf Experimental result reproducibility}
    \item[] Question: Does the paper fully disclose all the information needed to reproduce the main experimental results of the paper to the extent that it affects the main claims and/or conclusions of the paper (regardless of whether the code and data are provided or not)?
    \item[] Answer: \answerYes{}.
    \item[] Justification: Sections~\ref{sec:structured_tasks}, \ref{sec:open_ended_eval}, and \ref{sec:prompting_inference} define the tasks, metrics, prompting protocols, inference setup, parsing, and judging procedures; additional prompts and validation details are provided in the appendix.

\item {\bf Open access to data and code}
    \item[] Question: Does the paper provide open access to the data and code, with sufficient instructions to faithfully reproduce the main experimental results, as described in supplemental material?
    \item[] Answer: \answerYes{}.
    \item[] Justification: The paper is accompanied by benchmark annotations, evaluation scripts, prompt templates, and documentation sufficient to reproduce the reported benchmark metrics, subject to the access terms of the underlying audio sources and closed-source model APIs.

\item {\bf Experimental setting/details}
    \item[] Question: Does the paper specify all the training and test details (e.g., data splits, hyperparameters, how they were chosen, type of optimizer) necessary to understand the results?
    \item[] Answer: \answerYes{}.
    \item[] Justification: The dataset construction, splits, task definitions, model routing, prompting protocols, deterministic decoding policy, parsing, and judge validation are described in the paper and appendix.

\item {\bf Experiment statistical significance}
    \item[] Question: Does the paper report error bars suitably and correctly defined or other appropriate information about the statistical significance of the experiments?
    \item[] Answer: \answerYes{}.
    \item[] Justification: Comparisons with the majority baseline report paired percentile bootstrap 95\% intervals over 10,000 recording resamples (Appendix~\ref{app:category_analysis}), and supervised-control comparisons report paired bootstrap intervals over performer groups (Appendix~\ref{app:supervised_controls}); intervals are not adjusted for multiple comparisons. Other benchmark metrics are deterministic on fixed evaluation sets and are reported without error bars.

\item {\bf Experiments compute resources}
    \item[] Question: For each experiment, does the paper provide sufficient information on the computer resources (type of compute workers, memory, time of execution) needed to reproduce the experiments?
    \item[] Answer: \answerYes{}.
    \item[] Justification: Appendix~\ref{app:compute_resources} reports that the evaluated audio-language models are not trained or fine-tuned, the supervised controls train only small heads on frozen audio representations, open-weight inference uses a single NVIDIA H100 80GB GPU, closed-source models use public APIs, and long-running inference and judge jobs are executed with resumable batch scripts because runtime varies by model and API latency.
    
\item {\bf Code of ethics}
    \item[] Question: Does the research conducted in the paper conform, in every respect, with the NeurIPS Code of Ethics \url{https://neurips.cc/public/EthicsGuidelines}?
    \item[] Answer: \answerYes{}.
    \item[] Justification: We reviewed the NeurIPS Code of Ethics and believe the work conforms to it. The paper uses cited audio sources, expert annotation, documented licensing, and warns that model-generated vocal guidance should not replace professional instruction.

\item {\bf Broader impacts}
    \item[] Question: Does the paper discuss both potential positive societal impacts and negative societal impacts of the work performed?
    \item[] Answer: \answerYes{}.
    \item[] Justification: The introduction and conclusion discuss the positive goal of improving audio-language evaluation for expert feedback, while the limitations discuss risks from incorrect vocal guidance and the need not to treat model outputs as professional instruction.
    
\item {\bf Safeguards}
    \item[] Question: Does the paper describe safeguards that have been put in place for responsible release of data or models that have a high risk for misuse (e.g., pre-trained language models, image generators, or scraped datasets)?
    \item[] Answer: \answerNA{}.
    \item[] Justification: The paper does not release a generative model or other high-risk dual-use system. The benchmark release is limited to annotations, redistributable source audio, evaluation code, and documentation governed by the licenses and access terms of the underlying audio sources.

\item {\bf Licenses for existing assets}
    \item[] Question: Are the creators or original owners of assets (e.g., code, data, models), used in the paper, properly credited and are the license and terms of use explicitly mentioned and properly respected?
    \item[] Answer: \answerYes{}.
    \item[] Justification: Existing audio datasets and models are cited in the paper, and source-license information for the audio assets is documented in the appendix.

\item {\bf New assets}
    \item[] Question: Are new assets introduced in the paper well documented and is the documentation provided alongside the assets?
    \item[] Answer: \answerYes{}.
    \item[] Justification: VocalCoachBench introduces new expert annotations, prompts, evaluation code, and documentation; the paper and appendix describe the annotation schema, taxonomy, examples, validation, and intended benchmark use.

\item {\bf Crowdsourcing and research with human subjects}
    \item[] Question: For crowdsourcing experiments and research with human subjects, does the paper include the full text of instructions given to participants and screenshots, if applicable, as well as details about compensation (if any)? 
    \item[] Answer: \answerYes{}.
    \item[] Justification: The work does not use open crowdsourcing; annotations were collected from paid expert vocal coaches who consented to the annotation work. The paper describes the expert annotation protocol, quality-control process, annotation interface examples, and annotation labor details in the main text and Appendix.

\item {\bf Institutional review board (IRB) approvals or equivalent for research with human subjects}
    \item[] Question: Does the paper describe potential risks incurred by study participants, whether such risks were disclosed to the subjects, and whether Institutional Review Board (IRB) approvals (or an equivalent approval/review based on the requirements of your country or institution) were obtained?
    \item[] Answer: \answerNo{}.
    \item[] Justification: The work involved only paid annotation rather than a behavioral intervention. Under our applicable institutional policy, formal IRB review was not required. Annotators were informed of the task content and compensation before participation, and no private personal information was collected for analysis.

\item {\bf Declaration of LLM usage}
    \item[] Question: Does the paper describe the usage of LLMs if it is an important, original, or non-standard component of the core methods in this research? Note that if the LLM is used only for writing, editing, or formatting purposes and does \emph{not} impact the core methodology, scientific rigor, or originality of the research, declaration is not required.
    %this research? 
    \item[] Answer: \answerYes{}.
    \item[] Justification: LLMs are used as part of the methodology for claim decomposition, parsing/normalization, translation support, and judge-based evaluation; these uses are described in the dataset post-processing and evaluation sections, with the claim-extraction prompt and judge validation details in the appendix.

\end{enumerate}

\end{document}